\documentstyle[prb,aps,twocolumn]{revtex}
\begin{document}
\input epsf.tex
\wideabs{
\draft
\preprint{Submitted to PRB}
\title{Synthesis, Characterization and Magnetic Susceptibility of the Heavy
Fermion Transition Metal Oxide LiV$_2$O$_4$}
\author{S. Kondo, D.~C. Johnston, and L.~L.~Miller}
\address{Ames Laboratory and Department of Physics and Astronomy, Iowa State
University, Ames, Iowa 50011}
\date{Accepted for publication in Phys.\ Rev.\ B}

\maketitle

\begin{abstract} The preparative method, characterization and magnetic
susceptibility $\chi$ measurements versus temperature $T$ of the heavy fermion
transition metal oxide LiV$_2$O$_4$ are reported in detail.  The intrinsic
$\chi(T)$ shows a nearly $T$-independent behavior below $\sim 30$\,K with a
shallow broad maximum at $\approx 16$\,K, whereas Curie-Weiss-like behavior is
observed above $\sim 50$--100\,K\@.  Field-cooled and zero-field-cooled
magnetization $M^{\rm obs}$ measurements in applied magnetic fields
$H=10\mbox{--}100$\,G from 1.8 to 50\,K showed no evidence for spin-glass
ordering.  Crystalline electric field theory for an assumed cubic V point group
symmetry is found insufficient to describe the observed temperature variation of
the effective magnetic moment.   The Kondo and Coqblin-Schrieffer models do not
describe the magnitude and $T$ dependence of $\chi$ with realistic parameters.  In
the high $T$ range, fits of $\chi(T)$ by the predictions of high temperature 
series expansion calculations provide estimates of the V-V antiferromagnetic
exchange coupling constant $J/k_{\rm B}$ $\sim 20$\,K, $g$-factor $g \sim 2$ and
the $T$-independent susceptibility.  Other possible models to describe the
$\chi(T)$ are discussed.  The paramagnetic impurities in the samples were
characterized using isothermal $M^{\rm obs}(H)$ measurements with $0<H \leq
5.5$\,T at 2 to 6\,K\@.  These impurities are inferred to have spin
$S_{\rm imp}\sim 3/2$ to~4, $g_{\rm imp}\sim 2$ and molar concentrations of 0.01
to 0.8\%, depending on the sample.
\end{abstract}

}
\section{Introduction}\label{IntroSec}

Especially since the discoveries of heavy fermion (HF)\cite{Andres1975} and high
temperature superconducting compounds,\cite{Bednorz1986} strongly correlated
electron systems have drawn much attention both theoretically and experimentally. 
Extensive investigations have been done on many  cerium- and uranium-based HF
compounds.\cite{HFIV} The term ``heavy fermion'' refers to the large quasiparticle
effective mass $m^{*}/m_{\rm e} \sim 100\mbox{--}1000$ of these compounds inferred
from the electronic specific heat coefficient $\gamma(T) \equiv C_{\rm e}(T)/T$ at
low temperature $T$, where $m_{\rm e}$ is the  free electron mass and $C_{\rm e}$
is the electronic specific heat.  Fermi liquid (FL) theory explains well the
low-$T$ properties of many HF compounds.  Non-FL compounds\cite{Maple1994} are
currently under intensive study in relation to quantum critical
phenomena.\cite{ZulickeMillis1995}  The transition metal oxide compound
LiV$_2$O$_4$ was recently reported\cite{Kondo1997} to be the first $d$-electron
metal to show heavy FL behaviors characteristic of those of the heaviest mass
$f$-electron systems.

\begin{figure}
\epsfxsize=2in
\centerline{\epsfbox{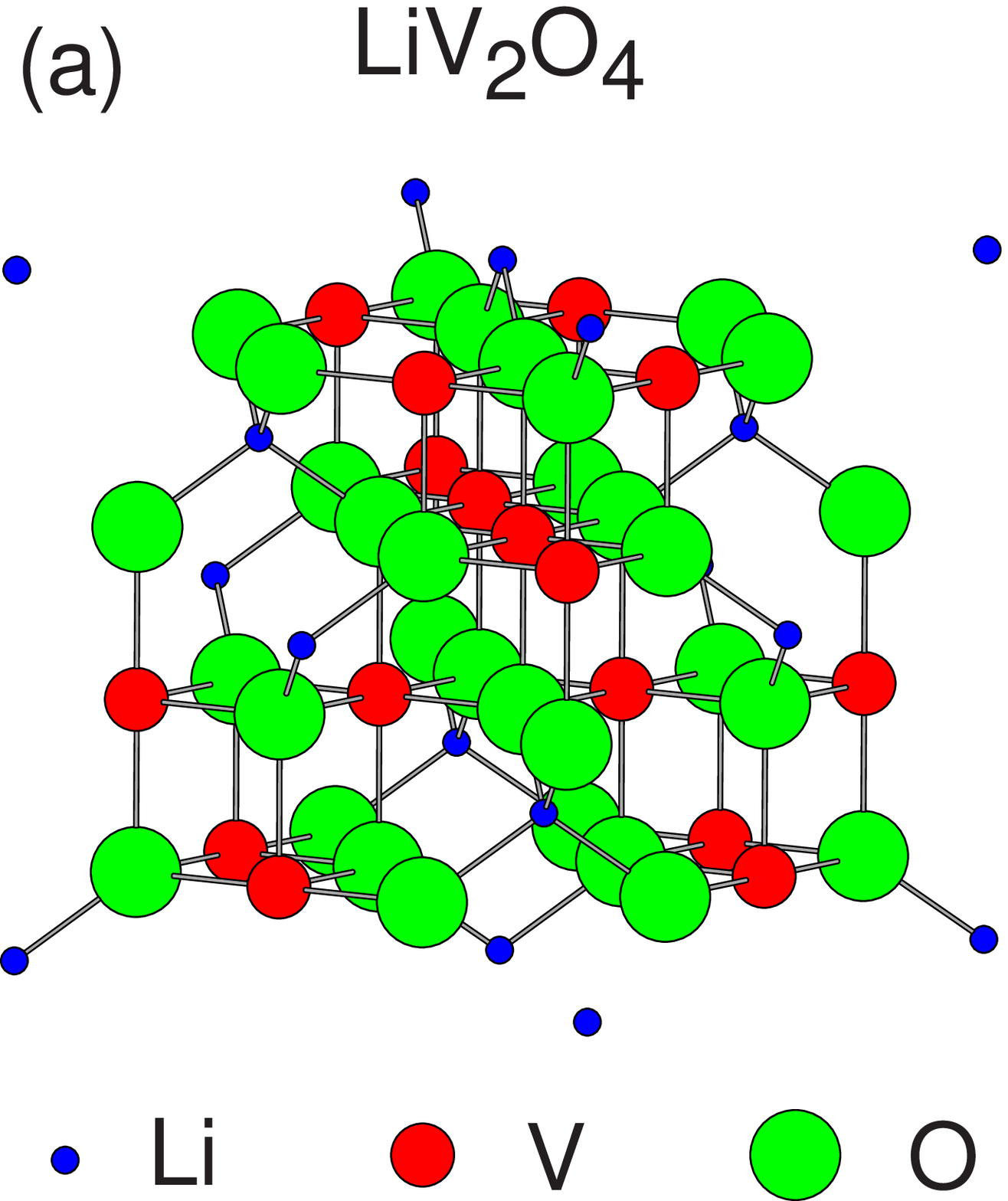}}
\vglue0.2in
\epsfxsize=1.3in
\centerline{\epsfbox{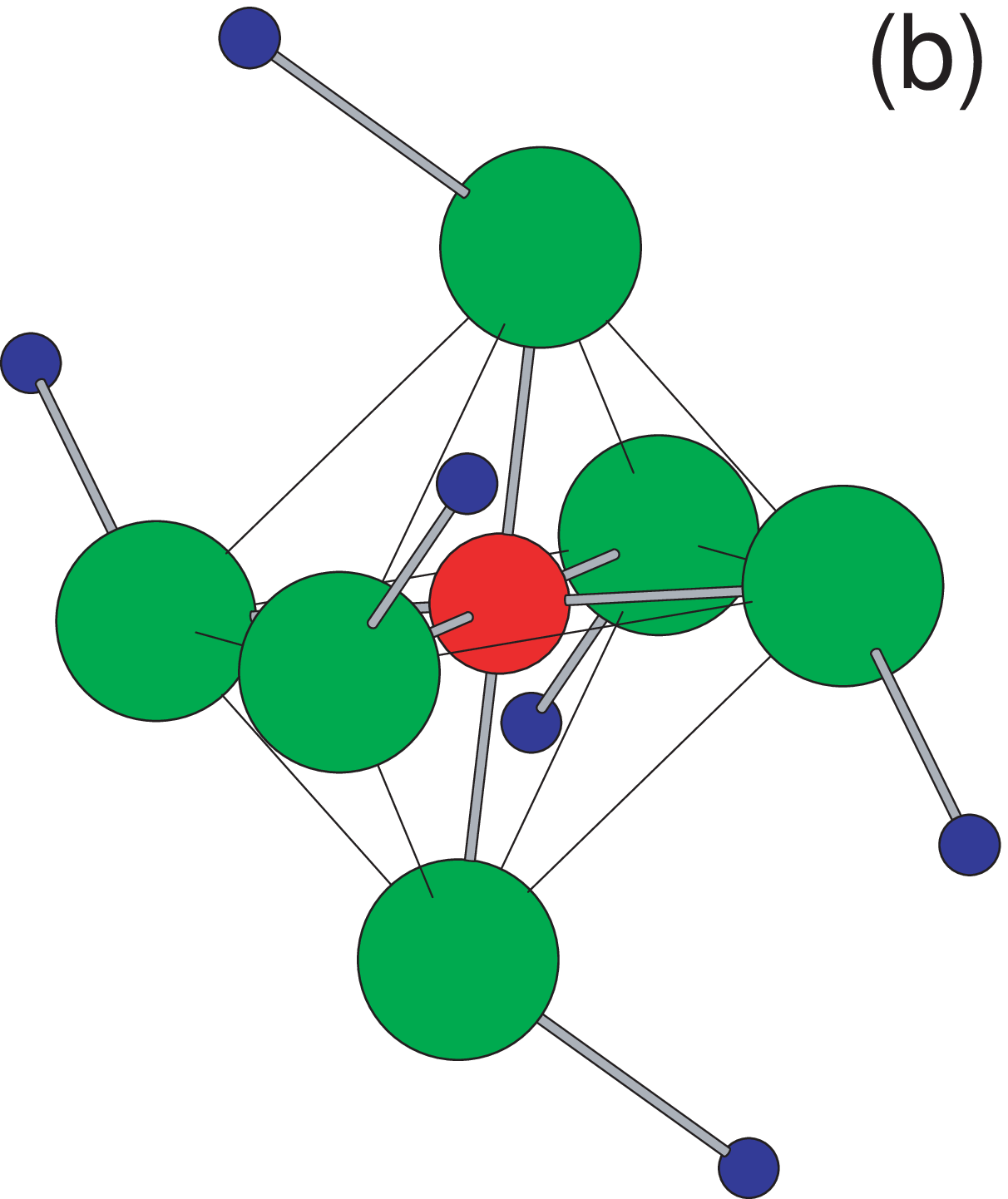}}
\vglue0.1in
\caption{(color) (a) Normal spinel structure of LiV$_2$O$_4$ with an fcc Bravais
unit cell.  (b) A part of the structure depicting the trigonally distorted oxygen 
octahedra.  The distortion shown is exaggerated for clarity and corresponds to an 
oxygen parameter $u=0.27$.  Small, medium and large spheres represent lithium,
vanadium and oxygen, respectively; their sizes have no intended physical
significance.}
\label{spinel}
\end{figure}

LiV$_2$O$_4$ has the  face-centered-cubic (fcc), normal-spinel structure with
space group $Fd\bar{3}m$ [Fig.~\ref{spinel}(a)], first synthesized by Reuter and
Jaskowsky in 1960.\cite{ReuterJaskowsky1960} The V ions have a formal oxidation
state of +3.5, assuming that those of Li and O are $+1$ and $-2$, respectively,
corresponding to 1.5 $d$-electrons per V ion.  In the normal oxide spinel
LiV$_2$O$_4$, the oxygen ions constitute a nearly cubic-close-packed array. 
Lithium occupies the $8a$ sites,\cite{IntlTable1987} corresponding to one-eighth
of the 64 tetrahedral holes formed by the close-packed oxygen sublattice in a
Bravais unit cell that contains eight Li[V$_2$]O$_4$ formula units.  Vanadium
occupies the $16d$ sites (enclosed in square brackets in the formula),
corresponding to one-
\noindent half of the 32 octahedral holes in the oxygen sublattice per
unit cell.  All of the V ions are crystallographically equivalent. Due to this
fact and the non-integral V oxidation state, the compound is expected to be
metallic, which was confirmed by single-crystal resistivity $\rho(T)$
measurements by Rogers {\em et~al}.\cite{Rogers1967}   The V atoms constitute a
three-dimensional network of corner-shared tetrahedra.  The LiV$_2$ sublattice is
identical to the cubic Laves phase (C15) structure, and the V sublattice is
identical with the transition metal $T$ sublattice of the fcc $R_2T_2$O$_7$
pyrochlore structure.

Despite its metallic character, LiV$_2$O$_4$ exhibits a strongly temperature
dependent magnetic susceptibility, indicating strong electron correlations. In the
work reported before 1997, the observed magnetic susceptibility $\chi^{\rm
obs}(T)$ was found to increase monotonically with decreasing $T$ down to
$\approx$~4\,K and to approximately follow the Curie-Weiss
law.
\cite{Kessler1971,Nakajima1991,ChamberlandHewston1986,Takagi1987,Hayakawa1989,%
Johnston1995}  
Kessler and Sienko\cite{Kessler1971} interpreted their $\chi^{\rm obs}(T)$ data as
the sum of a Curie-Weiss term $2C/(T-\theta)$ and a temperature-independent term
$\chi_0=0.4\times10^{-4}\,$cm$^3$/mol. Their Curie constant $C$ was
0.468\,cm$^3$\,K/(mol\,V), corresponding to a V$^{+4}$ $g$-factor of 2.23 with
spin $S=1/2$.  The negative Weiss temperature $\theta=-63$\,K suggests
antiferromagnetic (AF) interactions between the V spins.  However, no magnetic
ordering was found above $4.2$\,K\@.  This may be understood in terms of possible
suppression of long-range magnetic ordering due to the geometric frustration
among the AF-coupled V spins in the tetrahedra
network.\cite{Anderson1956,Villain1980}   Similar values of $C$ and $\theta$ have
also been obtained by subsequent workers,
\cite{Nakajima1991,ChamberlandHewston1986,Takagi1987,Hayakawa1989,Johnston1995}
as shown in Table~\ref{PastWorks}, in which reported crystallographic
data\cite{ReuterJaskowsky1966,Rogers1964,Pollert1973,Arndt1974,Ueda1997} are
also shown.  This local magnetic moment behavior of LiV$_2$O$_4$ is in marked
contrast to the magnetic properties of isostructural LiTi$_2$O$_4$ 
\widetext
\begin{table}
\caption{Lattice parameter $a_0$, oxygen parameter $u$ (see text) and magnetic
parameters $\chi_0$, $C$ and $\theta$ reported in the literature for
LiV$_2$O$_4$.  The $u$ values shown are for the second setting of the space
group $Fd\bar{3}m$ from the {\em International Tables for Crystallography, Vol.\
A.}\protect\cite{IntlTable1987}  The ``$T$ range'' is the temperature range over
which the fits to the susceptibility data were done, $\chi_0$ is the
temperature-independent contribution, $C$ is the Curie constant and $\theta$ is
the Weiss temperature.  The error in the last digit of a quantity is given in
parentheses.  Unless otherwise noted, all measurements were done on
polycrystalline samples.}
\label{PastWorks}
\begin{tabular}{ldccdcr}
$a_0$ & $u$ & $T$ range & $\chi_0$ & $C$ & $\theta$ & Ref.\\ (\AA)&&(K)&$(10^{-6}\
\frac{{\rm cm}^3}{{\rm mol\ LiV_2O_4}})$&$(\frac{{\rm cm^3\ K}}{{\rm mol\
V}})$&(K)&\\
\hline 8.22&&&&&&\protect\onlinecite{ReuterJaskowsky1960}\\
8.2403(12)&0.260(1)&&&&&\protect\onlinecite{ReuterJaskowsky1966}\\
8.240(2)&&&&&&\protect\onlinecite{Rogers1964}\\
8.22&&4.2--308&37&0.468&$-$63&\protect\onlinecite{Kessler1971}\\
8.240(2)&0.253(1)&&&&&\protect\onlinecite{Pollert1973}\\
8.25\tablenotemark[1]&&&&&&\protect\onlinecite{Arndt1974}\\
8.255(6)&0.260&50--380\tablenotemark[1]&37&0.460&$-$34&
\protect\onlinecite{ChamberlandHewston1986}\\
&&50--380\tablenotemark[1]&37&0.471&$-$42&
\protect\onlinecite{ChamberlandHewston1986}\tablenotemark[2]\\
&&80--300&43&0.441\tablenotemark[1]&$-$31\tablenotemark[1]
&\protect\onlinecite{Takagi1987}\\
8.241(3)\tablenotemark[1]&&80--300&43&0.434\tablenotemark[1]
&$-$39\tablenotemark[1]&\protect\onlinecite{Hayakawa1989}\\
&&&&0.473&&\protect\onlinecite{Nakajima1991}\\
8.235&&10--300&0&0.535&$-$35.4&\protect\onlinecite{Johnston1995}\\
8.2408(9)&&100--300&230&0.35&$-$33&\protect\onlinecite{Ueda1997}\\
\end{tabular}
\tablenotetext[1]{This value was digitized from the published figure.}
\tablenotetext[2]{Single crystal susceptibility data, corrected for the
contribution of 10\% V$_4$O$_7$.}
\end{table}
\narrowtext
\noindent
which
manifests a comparatively temperature independent Pauli paramagnetism and
superconductivity ($T_{\rm c}\leq 13.7$ K).\cite{Johnston1976}

Strong electron correlations in LiV$_2$O$_4$ were inferred by Fujimori {\em et
al}.\cite{Fujimori1988,Abbate1991} from their ultraviolet (UPS) and x-ray (XPS)
photoemission  spectroscopy measurements.  An anomalously small density of states
at the Fermi level was observed at room temperature which they attributed to the
effect of long-range Coulomb interactions.  They interpreted the observed spectra
assuming charge fluctuations between $d^1$ (V$^{4+}$) and $d^2$ (V$^{3+}$)
configurations on a time scale longer than that of photoemission ($\sim
10^{-15}$\,sec).  Moreover, the intra-atomic Coulomb repulsion energy, $U$, was
found to be $\sim 2$\,eV.  This value is close to the width $W \sim 2$\,eV of the
$t_{2g}$ conduction band calculated for
LiTi$_2$O$_4$.\cite{Satpathy1987,Massidda1988}  From these observations, one might
infer that $U \sim W$ for LiV$_2$O$_4$, suggesting proximity to a metal-insulator
transition.

We and collaborators recently reported that LiV$_2$O$_4$ samples with high
magnetic purity display a crossover from the aforementioned localized moment
behavior above $\sim 100$\,K to a nearly temperature independent susceptibility
below $\sim 30$\,K.\cite{Kondo1997}  This new finding was also reported
independently and nearly simultaneously by two other 
groups.\cite{Ueda1997,Onoda1997}  Specific heat measurements revealed a rapidly
increasing $\gamma(T)$ with decreasing temperature below $\sim 30$\,K with an
exceptionally large value $\gamma(1\,{\rm K}) \approx
0.42$\,J/mol\,K$^2$.\cite{Kondo1997}  To our knowledge, this $\gamma(1\,{\rm K})$
is the largest value reported for any metallic $d$-electron compound, {\em e.g.},
Y$_{0.97}$Sc$_{0.03}$Mn$_2$ ($\lesssim
0.2$\,J/mol\,~K$^2$)(Ref.~\onlinecite{Ballou1996}) and V$_{2-y}$O$_3$ ($\lesssim
0.07$\,~J/mol\,~K$^2$).\cite{Carter1993}  The Wilson ratio\cite{Wilson1975} at low
$T$ was found to be
$R_{\rm W}\sim$ 1.7, consistent with a heavy FL interpretation.  From $^7$Li NMR
measurements, the $T$ variation of the Knight shift $K$ was found to
approxi-
\newpage\noindent
mately follow that of the susceptibility.
\cite{Kondo1997,Onoda1997,Amako1990,Fujiwara1997,Fujiwara1998,Mahajan1998} 
The $^7$Li nuclear spin-lattice relaxation rate $1/T_1$ in LiV$_2$O$_4$ was  found
to be proportional to $T$ below $\sim 4$\,K, with a Korringa ratio on the order of
unity, again indicating FL
behavior.\cite{Kondo1997,Fujiwara1997,Fujiwara1998,Mahajan1998}

In this paper we present a detailed study of the synthesis, characterization and
magnetic susceptibility of LiV$_2$O$_4$.  In Sec.~\ref{ExpDetailSec} our synthesis
method and other experimental techniques are described.  Experimental results and
analyses are given in Sec.~\ref{ResultAnalysisSec}\@.  In Sec.~\ref{StructureSec},
after a brief overview of the spinel structure, we present structural
characterizations of nine LiV$_2$O$_4$ samples that were prepared in slightly
different ways, based upon our results of thermogravimetric analysis (TGA), x-ray
diffraction measurements and their Rietveld analyses.  In Sec.~\ref{MagnetismSec},
results and analyses of magnetization measurements are given.  In
Sec.~\ref{OverviewChiSec} an overview of the $\chi^{\rm obs}(T)\equiv M^{\rm
obs}(T)/H$ data of all nine samples studied is presented.  Then, in
Sec.~\ref{IsothermSec}, we determine the magnetic impurity concentrations from
analysis of the $M^{\rm obs}(H)$ data.  Low-field ($H=10\mbox{--}100$\,G)
$\chi^{\rm obs}(T)$ susceptibility data, measured after zero-field cooling (ZFC)
and field cooling (FC), are presented in
Sec.~\ref{SusceptibilitySec}~\ref{LowHChiSec}, from which we infer that spin-glass
ordering does not occur above 2\,K\@.  The above determinations of magnetic 
impurity contributions to $M^{\rm obs}(H,T)$ allow us to extract the intrinsic
susceptibility $\chi(T)$ from $\chi^{\rm obs}(T)$, as explained in
Sec.~\ref{SusceptibilitySec}~\ref{IntChiSec}.  The paramagnetic orbital Van
Vleck susceptibility $\chi^{\rm VV}$ contribution is determined in
Sec.~\ref{VVChiSec} from a so-called $K$-$\chi$ analysis using $^{51}$V NMR
measurements.\cite{Amako1990,Mahajan1998}  We attempt to interpret the $\chi(T)$
data using three theories.  First, the predictions of high temperature series
expansion (HTSE) calculations for the spin $S=1/2$ Heisenberg model are compared
to our $\chi(T)$ data in Sec.~\ref{HTSESec}.  Second, a crystalline electric
field theory prediction with the assumption of cubic point symmetry of the
vanadium ion is tested in Sec.~\ref{CFTSec}.  Third, we test the applicability of
the Kondo and Coqblin-Schrieffer models to our $\chi(T)$ data in
Sec.~\ref{KondoCSModelsSec}\@.  A summary and discussion are given in
Sec.~\ref{ConclusionSec}\@.  Throughout this paper, a ``mol'' means a mole 
\widetext
\begin{table}
\caption{Results of Rietveld refinements of x-ray diffraction measurements and
magnetization $M^{\rm obs}(H)$ isotherm analyses. The oxygen parameter ($u$) is 
for the second setting of the space group $Fd\bar{3}m$ from the {\it
International Tables for Crystallography, Vol.\ A}.\protect\cite{IntlTable1987} 
$f_{\rm str~imp}$ is the impurity concentration.  The error in the last digit of a
quantity is given in parentheses.  The detection limit of $f_{\rm str~imp}$ is
assumed to be 1\%.\protect\cite{Izumi1998}  For samples 3 and 7 in which no
discernable impurities were seen, this detection limit is listed; the Rietveld
refinement for sample 5 directly yielded  $f_{\rm str~imp}<1$\%.}
\label{Riet-Results}
\begin{tabular}{ccccccc} Sample & Alt. Sample & Cooling & Impurity & $a_0$ & $u$ &
$f_{\rm str~imp}$\\  No. & No. & & & (\AA) & & (mol\,\%) \\ \hline 1 & 4-0-1 & air
& V$_3$O$_5$ & 8.24062(11) & 0.26115(17) & 2.01\\ 2 & 3-3 & air & V$_2$O$_3$ &
8.23997(4) & 0.2612(20) & 1.83\\ 3 & 4-E-2 & air & pure & 8.24100(15) &
0.26032(99) & $<1$\\ 4 & 3-3-q1 & LN$_2$ & V$_3$O$_5$ & 8.24622(23) & 0.26179(36)
& 3.83\\ 4A & 3-3-q2 & ice H$_2$O & V$_2$O$_3$ & 8.24705(29) & 0.26198(39) &
1.71\\ 4B & 3-3-a2 & slow cool & V$_2$O$_3$ & 8.24734(20) & 0.26106(32) & 1.46\\ 5
& 6-1 & air & V$_2$O$_3$ & 8.24347(25) & 0.26149(39) & $<1$\\ 6 & 12-1 & air &
V$_3$O$_5$ & 8.23854(11) & 0.26087(23) & 2.20\\ 7 & 13-1 & air & pure & 8.24114(9)
& 0.26182(19) & $<1$\\
\end{tabular}
\end{table}
\noindent of
LiV$_2$O$_4$ formula units, unless otherwise noted.

\narrowtext
\section{Synthesis and Experimental Details}\label{ExpDetailSec}

Polycrystalline samples of LiV$_2$O$_4$ were prepared using conventional
solid-state reaction techniques with two slightly different paths to the
products.  The five samples used in our previous work\cite{Kondo1997} (samples 1
through 5) were prepared by the method in Ref.~\onlinecite{Johnston1976}.  Two
additional samples (samples 6 and 7) were synthesized by the method of Ueda {\em
et al}.\cite{Ueda1997}  Different precursors are used in the two methods:
``Li$_2$VO$_{3.5}$'' (see below) and Li$_3$VO$_4$, respectively. Both methods
successfully yielded high quality LiV$_2$O$_4$ samples which showed the broad peak
in $\chi^{\rm obs}(T)$ at $\approx 16$\,K\@.  In this report, only the first
synthesis method is explained in detail, and the reader is referred to
Ref.~\onlinecite{Ueda1997} for details of the second method.

The starting materials were Li$_2$CO$_3$ (99.999\,\%, Johnson Matthey),
V$_2$O$_3$, and V$_2$O$_5$ (99.995\,\%, Johnson Matthey).  Oxygen vacancies tend
to be present in commercially obtained V$_2$O$_5$.\cite{Ueda1}  Therefore, the
V$_2$O$_5$ was heated in an oxygen stream at 500-550\,$^{\circ}$C in order to
fully oxidize and also dry it.  V$_2$O$_3$ was made by reduction of either
V$_2$O$_5$ or NH$_4$VO$_3$ (99.995\,\%, Johnson Matthey) in a tube furnace under
5\,\%\,H$_2$/95\,\%\,He gas flow.  The heating was done in two steps: at
635\,$^\circ$C for $\approx 1$\, day and then at 900--1000\,$^\circ$C for up to 3
days.  The oxygen content of the nominal V$_{2-y}$O$_3$ obtained was then
determined by thermogravimeteric analysis (TGA, see below). The precursor
``Li$_2$VO$_{3.5}$" (found to be a mixture of Li$_3$VO$_4$ and LiVO$_3$ from an
x-ray diffraction measurement) was prepared by heating a mixture of Li$_2$CO$_3$
and V$_2$O$_5$ in a tube furnace under an oxygen stream at $\approx
525$\,$^\circ$C until the expected weight decrease due to the loss of carbon
dioxide was obtained.  Ideally the molar ratio of Li$_2$CO$_3$ to V$_2$O$_5$ for
the nominal composition Li$_2$VO$_{3.5}$ is 2 to 1.  A slight adjustment was,
however, made to this ratio according to the actual measured oxygen content of the
V$_{2-y}$O$_3$ ($y\simeq 0.005$ to 0.017)  so that the final product
\newpage\noindent
is stoichiometric LiV$_2$O$_4$.   This precursor and V$_{2-y}$O$_3$ were ground
thoroughly inside a helium-filled glovebox.  The mixture was then pelletized,
wrapped in a piece of gold foil, sealed into a quartz tube under vacuum, and
heated between 570\,$^\circ$C and 700\,$^\circ$C for $\lesssim 2$\, weeks.  The
as-prepared samples were all  removed from the oven at the final furnace
temperature and air-cooled to room temperature.  For samples 2 and 3  additional
heating at a higher
$T=750$\,$^\circ$C was given, with a repeated sequence of grinding, repelletizing
and reheating for sample 2.  From $\approx 725$\,$^\circ$C different methods of
cooling, liquid-nitrogen or ice-water quenching or slow-oven cooling, were applied
to pieces from sample 2, yielding samples 4, 4A and 4B, respectively.

Using a Rigaku Geigerflex diffractometer with a curved graphite crystal
monochrometer, x-ray diffraction patterns were obtained at room temperature with
Cu K$\alpha$ radiation.  \mbox{Rietveld} analyses of the diffraction patterns
were carried out using the angle-dispersive x-ray diffraction version of the
RIETAN-97$\beta$ program.\cite{Izumi1993,Izumi1998}

TGA measurements were done using a Perkin-Elmer TGA 7 Thermogravimetric Analyzer. 
Oxygen contents of the samples were calculated from weight gains after heating in
an oxygen flow to 540\,$^{\circ}$C for LiV$_2$O$_4$ and 620\,$^{\circ}$C for
V$_{2-y}$O$_3$, assuming that the oxidized products contained vanadium as V$^{+5}$.

Magnetization $M^{\rm obs}$ measurements were performed using a Quantum Design
MPMS5 superconducting quantum interference device (SQUID) magnetometer over the
$T$ range from 1.8--2\,K to 400\,K with $H$ up to 5.5\,T.  Zero-field-cooled (ZFC,
usually obtained by quenching the superconducting solenoid)
$M^{\rm obs}(H=1\,{\rm T},T)$ scans were carried out and isothermal $M^{\rm
obs}(H)$ data at various temperatures were obtained.  Low-field (10--100\,G) ZFC
and field-cooled (FC) $M(T)$ scans were done from 1.8--2\,K to 50\,K in order to
check for the presence or absence of spin-glass ordering.

\section{Results and Analyses}\label{ResultAnalysisSec}

\subsection{Structure}\label{StructureSec}

X-ray diffraction patterns of our nine LiV$_2$O$_4$ samples revealed that the
samples were single-phase or very nearly so.  Figure \ref{XraySamples1, 2 and
7}(a) shows the diffraction pattern of sample 7 which has no detectable
impurities.  The nine samples described in detail in this paper are categorized
into three groups in terms of purity: essentially impurity-free (samples 3 and 7),
V$_3$O$_5$ impurity (samples 1, 4 and 6) and V$_2$O$_3$ impurity (samples 2, 4A,
4B and 5).  The presence of these impurity phases is detected in magnified views
of the diffraction patterns as shown in Fig.~\ref{XraySamples1, 2 and 7}(b). Results from Rietveld analyses of the diffraction patterns for these samples are
given in Table~\ref{Riet-Results}. The refinements of the spinel phase (space
group $Fd\bar{3}m$, No.~227) were based on the assumption of exact LiV$_2$O$_4$
stoichiometry and the normal-spinel structure cation distribution.  The values of
the isotropic thermal-displacement parameters $B$ of lithium and oxygen were taken
from the Rietveld analysis of neutron diffraction measurements on our LiV$_2$O$_4$
sample 5 by Chmaissem {\em et al.},\cite{Chmaissem1997} and fixed throughout to
$B_{\rm Li}=1.1$\,\AA\ and
$B_{\rm O}=0.48$\,\AA, respectively.  These two atoms do not scatter x-rays
strongly enough to allow  accurate determinations of the $B$ values from Rietveld
refinements of our x-ray diffraction data.
\begin{figure}
\epsfxsize=3.1in
\centerline{\epsfbox{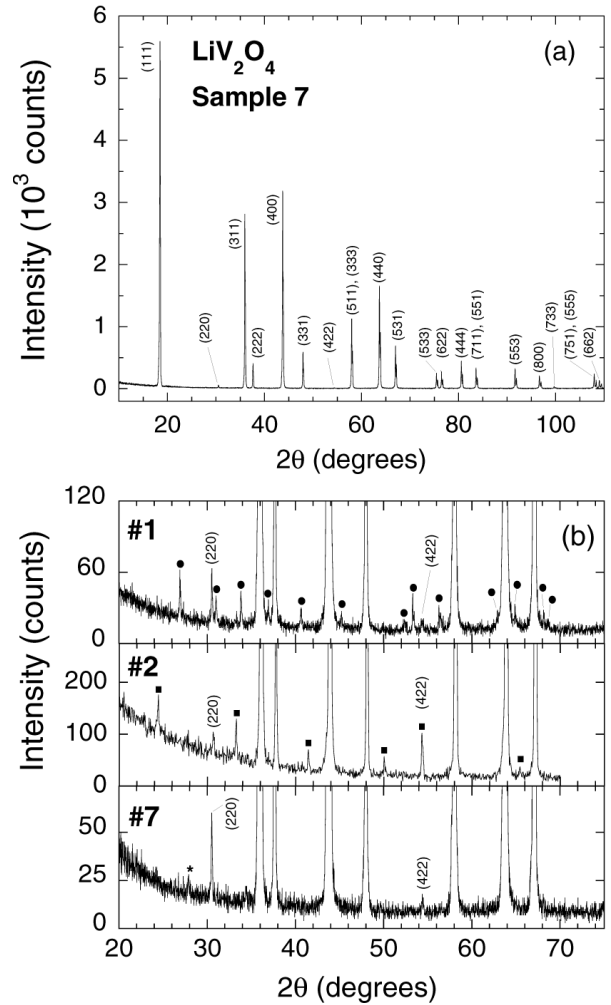}}
\vglue 0.1in
\caption{(a) X-ray diffraction pattern of LiV$_2$O$_4$ sample 7.  The spinel-phase
peaks are indexed as shown.  (b) Expanded plots of the X-ray patterns of samples 1
(top), 2 (middle) and 7 (bottom).  Indexed peaks are those of the spinel phase. 
Sample 1 has V$_3$O$_5$ impurity (filled circles), whereas sample 2 has V$_2$O$_3$
impurity (filled squares).  Sample 7 has no impurity peaks except possibly the
very weak unidentified one marked with a star.}
\label{XraySamples1, 2 and 7}
\end{figure}

The positions of the oxygen atoms within the unit cell of the spinel structure are
described by a variable oxygen parameter $u$ associated with the $32e$ positions
in space group $Fd\bar{3}m$.  The value of $u$ [in the space group setting with
the origin at center ($\bar{3}m$)] for each of our samples was found to be larger
than the ideal close-packed-oxygen value of 1/4.  Compared to the ``ideal''
structure with $u=1/4$, the volumes of an oxygen tetrahedron and an octahedron
become larger and smaller, respectively.  The increase of the tetrahedron volume
takes place in such a way that each of the four Li-O bonds are lengthened along
one of the $<$111$>$ directions, so that the tetrahedron remains undistorted.  As
a result of this elongation, the tetrahedral and octahedral holes become
respectively larger and smaller.\cite{Blasse1964}  Each of the oxygen atoms in a
tetrahedron is also bonded to three V atoms. Since the fractional coordinates of
both Li and V are  fixed in terms of the unit cell edge, an oxygen octahedron
centered by a V atom is accordingly trigonally distorted.  This distortion is
illustrated in Fig.~\ref{spinel}(b).

The nine LiV$_2$O$_4$ samples were given three different heat treatments after
heating to 700 to 750\,$^\circ$C: air-cooling (samples 1, 2, 3, 5, 6 and 7),
liquid-nitrogen quenching (sample 4), ice-water quenching (sample 4A) or oven-slow
cooling at $\approx 20\,^{\circ}$C/hr (sample 4B).  Possible loss of Li at the
high synthesis  temperature, perhaps in the form of a lithium oxide, was a
concern.  In a detailed neutron diffraction study, Dalton {\em
et~al.}\cite{Dalton1994}  determined the lithium contents in their samples of
Li$_{1+x}$Ti$_{2-x}$O$_4$ ($0 \leq x \leq 0.33$), and found lithium deficiency in
the $8a$ site of the spinel phase of all four samples studied.  If the spinel
phase in the Li-V-O system is similarly Li-deficient, then samples of exact
stoichiometry LiV$_2$O$_4$ would contain V-O impurity phase(s), which might then
explain the presence of small amounts of V$_2$O$_3$ or V$_3$O$_5$ impurity phases
in most of our samples.

\mbox{Sample~3~was intentionally made slightly off-} stoichiometric, with the
nominal composition LiV$_{1.92}$O$_{3.89}$.  A TGA measurement in oxygen showed a
weight gain of 12.804\,\% to the maximally oxidized state.  If one assumes an
actual initial composition LiV$_{1.92}$O$_{3.89+\delta}$, this weight gain
corresponds to
$\delta=0.08$ and an actual initial composition of LiV$_{1.92}$O$_{3.97}$ which
can be rewritten as Li$_{1.01}$V$_{1.93}$O$_4$ assuming no oxygen vacancies on the
oxygen sublattice.  On the other hand, if one assumes an actual initial
composition of Li$_{1-x}$V$_{1.92}$O$_{3.89}$, then the weight gain yields
$x=0.19$, and an initial composition Li$_{0.81}$V$_{1.92}$O$_{3.89}$ which can be
similarly rewritten as Li$_{0.83}$V$_{1.97}$O$_4$.  Our Rietveld refinements could
not distinguish these possibilities from the stoichiometric composition
Li[V$_2$]O$_4$ for the spinel phase.

Sample 4, which was given a liquid-nitrogen quench from the final heating
temperature of $\simeq 725\,^{\circ}$C (labelled ``LN$_2$'' in
Table~\ref{Riet-Results}), is one of the structurally least pure samples (see
Table~\ref{Riet-Results}).  Our Rietveld refinement of the x-ray diffraction
pattern for this sample did not reveal any discernable deviation of the cation
occupancy from that of ideal Li[V$_2$]O$_4$.  There is a strong similarity among
samples 4, 4A (ice-water quenched) and 4B (oven-slow cooled), despite their
different heat treatments.  These samples all have much larger lattice parameters
($a_{0} \gtrsim 8.246$\,\AA) than the other samples.  The as-prepared sample 2,
from which all three samples 4, 4A and 4B were obtained by the above quenching
heat treatments, has a much smaller lattice parameter.  On the other hand, the
oxygen parameters $u$ of these four samples are similar to each other and to those
of the other samples in Table~\ref{Riet-Results}.

The weight gains on oxidizing our samples in oxygen in the TGA can be converted to
values of the average oxidation state per vanadium atom, assuming the ideal
stoichiometry LiV$_2$O$_4$ for the initial composition.  The values, to an
accuracy of $\pm 0.01$, are 3.57, 3.55, 3.60, 3.56, 3.56, 3.57, 3.57, 3.55 for
samples 1--7 and 4B, respectively.  This measurement was not done for sample
4A\@.  These values are systematically higher than the expected value of 3.50,
possibly because the samples were not completely oxidized.  Indeed, the oxidized
products were gray-black, and upon crushing were brown, rather than a light
color.  On the other hand, x-ray diffraction patterns of the ``LiV$_2$O$_{5.5}$''
oxidation products showed only a mixture of LiVO$_3$ and Li$_4$V$_{10}$O$_{27}$
phases as expected from the known Li$_2$O-V$_2$O$_5$ phase
diagram.\cite{Reisman1962}  Our upper temperature limit (540$^{\circ}$C) during
oxidation of the LiV$_2$O$_4$ samples was chosen to be low enough so that the
oxidized product at that temperature contained no liquid phase; this temperature
may have been too low for complete oxidation to occur.  In contrast, our
V$_{2-y}$O$_3$ starting materials turned orange on oxidation, which is the same
color as the V$_2$O$_5$ from which they were made by hydrogen reduction.

\subsection{Magnetization Measurements}\label{MagnetismSec}

\subsubsection{Overview of Observed Magnetic Susceptibility}\label{OverviewChiSec}
\begin{figure}
\epsfxsize=3in
\centerline{\epsfbox{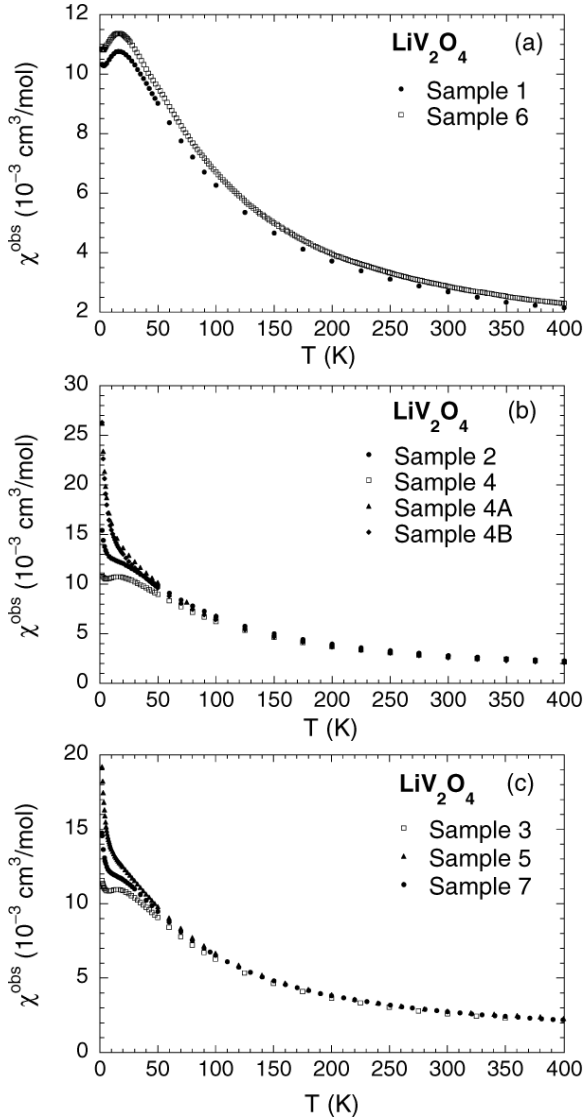}}
\vglue 0.1in
\caption{\mbox{Observed magnetic susceptibility $\chi^{\rm obs}(T)$} ($\equiv
M^{\rm obs}/H$) of all the nine samples studied, measured with $H=1$\,T after
being zero-field cooled to the lowest $T$: (a)~Samples 1 and 6; (b)~samples 2, 4,
4A and 4B; (c)~samples 3, 5 and 7.}
\label{ObsChiFigs}
\end{figure}

An overview of the observed ZFC magnetic susceptibilities $\chi^{\rm obs}(T)
\equiv M^{\rm obs}(T)/H$ at $H=1.0$\,T from 1.8--2\,K to 400\,K of the nine
LiV$_2$O$_4$ samples is shown in Figs.~\ref{ObsChiFigs} (a), (b) and (c).  The
$\chi^{\rm obs}(T)$ data for the various samples show very similar
Curie-Weiss-like behavior for $T \gtrsim 50$\,K\@. Differences in $\chi^{\rm
obs}(T)$ between the samples appear at lower $T$, where variable Curie-like
$C_{\rm imp}/T$ upturns occur.

Samples 1 and 6 clearly exhibit shallow broad peaks in $\chi^{\rm obs}$ at $T
\approx 16$\,K\@.  The
$\chi^{\rm obs}(T)$ of sample 6 is systematically slightly larger than that of
sample 1; the reason for this shift is not known.  Samples 3 and 4 also show the
broad peak with a relatively small Curie-like upturn.  Samples 2 and 7 show some
evidence of the broad peak but the peak is partially masked by the upturn.  For
samples~4A, 4B and 5, the broad peak is evidently masked by larger Curie impurity
contributions.  From Fig.~\ref{ObsChiFigs} and Table~\ref{Riet-Results}, the
samples 1, 4 and 6 with the smallest Curie-like magnetic impurity contributions
contain V$_3$O$_5$ impurities, whereas the other samples, with larger magnetic
impurity contributions, contain V$_2$O$_3$ impurities.  The reason for this
correlation is not clear.  The presence of the vanadium oxide impurities by itself
should not be a direct cause of the Curie-like upturns.  The susceptibility of
pure V$_2$O$_3$ follows the Curie-Weiss law in the metallic $T$ region above $\sim
170$\,K, but for $T\lesssim 170$\,K it becomes an antiferromagnetic insulator,
showing a decrease in 
\noindent
$\chi(T)$.\cite{Ueda1995}  V$_{2-y}$O$_3$ ($y\approx 0.03$), on the other hand,
sustains its high-$T$ metallic state down to low temperatures, and at its
N\'{e}el temperature $T_{\rm N}\sim 10$\,K it undergoes a transition to an
antiferromagnetic phase with a cusp in $\chi(T)$.\cite{Ueda1995}  V$_3$O$_5$ also
orders antiferromagnetically at $T_{\rm N}=75.5$\,K, but $\chi(T)$ shows a broad
maximum at a higher $T=125$\,K.\cite{Nagata1979}  Though not detected in our x-ray
diffraction measurements, V$_4$O$_7$, which has the same V oxidation state as in
LiV$_2$O$_4$, also displays a cusp in $\chi^{\rm obs}(T)$ at $T_{\rm N}\approx
33$\,K and $\chi^{\rm obs}(T)$ follows the Curie-Weiss law for $T
\gtrsim 50$\,K\@.\cite{Nagata1979}  The susceptibilities of these V-O phases are
all on the order of $10^{-4}$ to $10^{-3}$\,cm$^3$/mol at low
$T$.\cite{Ueda1995,Nagata1979}  Moreover, the $T$ variations of $\chi^{\rm
obs}(T)$ in these vanadium oxides for $T\lesssim 10$\,K are, upon decreasing $T$,
decreasing (V$_{2-y}$O$_3$) or nearly $T$ independent (V$_3$O$_5$ and V$_4$O$_7$),
in contrast to the increasing behavior of our Curie-like impurity
susceptibilities.  From the above discussion and the very small amounts of V-O
impurity phases found from the Rietveld refinements of our x-ray diffraction
measurements, we conclude that the V-O impurity phases cannot give rise to the
observed Curie-like upturns in our $\chi^{\rm obs}(T)$ data at low $T$.  These
Curie-like terms therefore most likely arise from paramagnetic defects in the
spinel phase and/or from a very small concentration of an unobserved impurity
phase.

Figure~\ref{ObsChiFigs}(b) shows how the additional heat treatments of the
as-prepared sample~2 yield different behaviors of $\chi^{\rm obs}(T)$ at low
$T$ in samples 4, 4A and 4B\@.  Only liquid-nitrogen quenching (sample 4) caused a
decrease in the Curie-like upturn of sample~2.  On the contrary, ice water
quenching (sample 4A) and oven-slow cooling (sample 4B) caused $\chi^{\rm obs}(T)$
to have an even larger upturn.  However, the size of the Curie-like upturn in
$\chi^{\rm obs}(T)$ of sample 4 was found to be irreproducible when the same
liquid-nitrogen quenching procedure was applied to another piece from sample~2; in
this case the Curie-like upturn was larger, not smaller, than in sample~2.  The
observed susceptibility (not shown) of this latter liquid nitrogen-quenched sample
is very similar to those of samples 4A and 4B.  The $\chi^{\rm obs}(T)$ of samples
4A and 4B resemble those reported
previously.
\cite{Kessler1971,Nakajima1991,ChamberlandHewston1986,Takagi1987,Hayakawa1989,%
Johnston1995}

\subsubsection{Isothermal Magnetization versus Magnetic Field}\label{IsothermSec}

\begin{figure}
\epsfxsize=3in
\centerline{\epsfbox{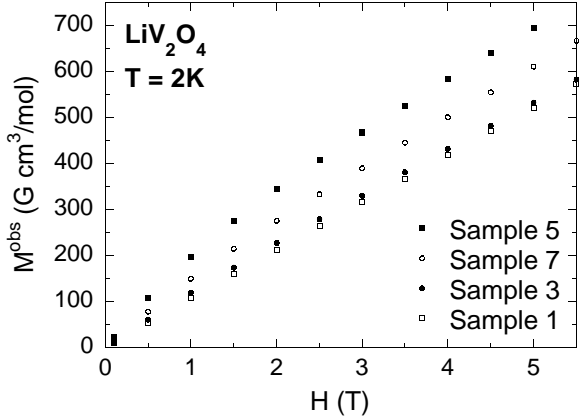}}
\vglue 0.1in
\caption{Comparison of the negative curvatures of observed magnetization isotherms
$M^{\rm obs}$ at
$T=2$\,K versus applied magnetic field $H$ for samples 1, 3, 5 and 7.}
\label{ObsMComparisonFig}
\end{figure}

Larger Curie-like upturns were found in samples with larger curvatures in the
isothermal $M^{\rm obs}(H)$ data at low
$T$.  A few representative $M^{\rm obs}(H,2\,{\rm K})$ data for samples showing
various extents of curvatures in $M^{\rm obs}(H)$ are shown in
Fig.~\ref{ObsMComparisonFig}, which may be compared with the corresponding
$\chi^{\rm obs}(T)$ data at low $T$ in Figs.~\ref{ObsChiFigs}.  This correlation
suggests that the Curie-like upturns in $\chi^{\rm obs}(T)$ arise from
paramagnetic (field-saturable) impurities/defects in the samples.  On the other
hand, there is no obvious correlation between the magnetic impurity concentration
and the V$_2$O$_3$ or V$_3$O$_5$ phase impurity concentration, as noted above.

\begin{figure}
\epsfxsize=3.1in
\centerline{\epsfbox{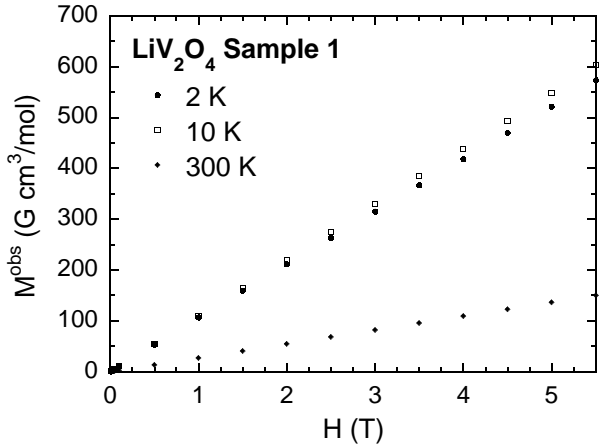}}
\vglue 0.1in
\caption{Observed magnetization $M^{\rm obs}$ versus applied magnetic field $H$
isotherms at temperatures $T=$ 2, 10 and 300\,K for LiV$_2$O$_4$ sample 1. 
Negative curvature in $M^{\rm obs}(H)$ is not present for $T > 10$\,K for this
sample.}
\label{IsothermSample1}
\end{figure}

The isothermal $M^{\rm obs}(H)$ data for $H \leq 5.5$\,T displayed negative
curvature for $T \lesssim 10$--20\,K and linear behavior for higher $T$, as
illustrated for sample 1 in Fig.~\ref{IsothermSample1}.  The concentrations and
other parameters of the magnetic impurities in the various samples were obtained
from analyses of $M^{\rm obs}(H)$ isotherms as follows.  From high-field
measurements, the intrinsic magnetization $M(H,0.5\,{\rm K})$ of LiV$_2$O$_4$ is
proportional to $H$ up to $H\sim$ 16\,T.\cite{Lacerda1998} Therefore, the observed
molar magnetization $M^{\rm obs}(H,T)$ isotherm data for each sample were fitted
by the equation
\begin{eqnarray} &M^{\rm obs}(H, T)\ & =  M_{\rm imp}(H, T) + M(H, T)\nonumber \\ 
 &&=  f_{\rm imp}N_{\rm A}g_{\rm imp}\mu_{\rm B}S_{\rm imp}B_{S_{\rm imp}}(x)
+\chi(T) H~,\nonumber\\
\label{MEq}
\end{eqnarray}
\noindent
where $f_{\rm imp}$ is the magnetic impurity concentration, $N_{\rm A}$
Avogadro's number, $g_{\rm imp}$ the impurity $g$-factor, $\mu_{\rm B}$ the Bohr
magneton,
$S_{\rm imp}$ the impurity spin, $B_{S_{\rm imp}}$ the Brillouin function, $\chi$
the intrinsic susceptibility of the LiV$_2$O$_4$ spinel phase and $H$ the applied
magnetic field.  The argument of the Brillouin function is $x=g_{\rm imp}\mu_{\rm
B}S_{\rm imp}H/[k_{\rm B}(T-\theta_{\rm imp})]$.  $\theta_{\rm imp}$ represents
the Weiss temperature of the Curie-Weiss law when the susceptibility is obtained
by expanding the Brillouin function in the limit of small $H/(T-\theta_{\rm
imp})$.  Incorporating the parameter $\theta_{\rm imp}\neq 0$ takes account of
possible interactions between magnetic impurities in a mean-field manner.  To
improve the precision of the obtained fitting parameters, we fitted $M^{\rm
obs}(H)$ isotherm data measured at more than one low temperature simultaneously by
Eq.~(\ref{MEq}).  Since the negative curvature of the isothermal $M^{\rm
obs}(H,T)$ data diminishes rapidly with increasing $T$, only low
$T$ (1.8--6\,K) data were used.  Furthermore, a linear $T$ dependence of $\chi(T)$
in this $T$ range was assumed [see Fig.~\ref{ObsChiFigs}(a)] in order to reduce
the number of free parameters.  However,
$\chi(T=2\,{\rm K})$ and the linear slope d$\chi$/d$T$ still have to be
determined.  Hence up to six free parameters were to be determined by fitting
Eq.~(\ref{MEq}) to the data: $f_{\rm imp}$, $g_{\rm imp}$, $S_{\rm imp}$,
$\theta_{\rm imp}$, $\chi(T=2\,{\rm K})$ and d$\chi$/d$T$.

With all six parameters varied as free parameters, fits of $M^{\rm obs}(H,T)$ by
Eq.~(\ref{MEq}) produced unsatisfactory results, yielding parameters with very
large estimated standard deviations.  Therefore, we fixed $S_{\rm imp}$ to various
half-integer values starting from 1/2, thereby reducing the number of free
parameters of each fit to five.  With regard to the $g_{\rm imp}$ values,
$g$-factors of slightly less than 2 are observed in V$^{+4}$ compounds: VO$_2$
(1.964) (Ref.~\onlinecite{Belyakov1973}), (NH$_4$)$_x$V$_2$O$_5$ (1.962)
(Ref.~\onlinecite{Palanisamy1974}) and Li$_x$V$_2$O$_5$ (1.96).\cite{Gendell1962} 
Using $g_{\rm imp} \approx 2$ as a guide, we selected a few values of $S_{\rm
imp}$ which resulted in $g\sim 2$ in the five-parameter fit.  Then using the
obtained parameter values we calculated and plotted the impurity magnetization
$M_{\rm imp}$ ($\equiv M^{\rm obs}-\chi H$) versus $H/(T-\theta_{\rm imp})$ for
all the low $T$ data utilized in the fit by Eq.~(\ref{MEq}).  If a fit is valid,
then all the $M_{\rm imp}[H/(T-\theta_{\rm imp})]$ data points obtained at the
various isothermal temperatures for each sample should collapse onto a universal
curve described by $M_{\rm imp}=f_{\rm imp}N_{\rm A}g_{\rm imp}\mu_{\rm B}S_{\rm
imp}B_{S_{\rm imp}}(x)$.  The fixed value of $S_{\rm imp}$ which gave the best
universal behavior for a given sample was chosen.  Then, using this $S_{\rm imp}$,
we fixed the value of $g_{\rm imp}$ to 2 to see if the resultant
$M_{\rm imp}[H/(T-\theta_{\rm imp})]$ data yielded a similar universal behavior. 
For the purpose of reducing the number of free parameters as much as possible, if
this fixed-$g$ fit did yield a comparable result, the parameters obtained were
taken as the final fitting parameters and are reported in this paper.  For sample
1 only, the fit parameters obtained by further fixing $\theta_{\rm imp}=0$ are
reported here.  To estimate the goodness of a fit, the $\chi^2$ per degree of
freedom (DOF) was obtained, which is defined as $(N_{\rm
p}-P)^{-1}\sum_{i=1}^{N_{\rm p}}(M_{i}-M_{i}^{\rm calc})^{2}/\sigma_{i}^{2}$,
where $N_{p}$ is the number of data points, $P$ is the number of free parameters,
and $\sigma_{i}$ is the standard deviation of the observed value $M_{i}$.  A fit
is regarded as satisfactory if $\chi^{2}/{\rm DOF} \lesssim 1$, and this criterion
was achieved for each of the nine samples.

\begin{figure}
\epsfxsize=3.1in
\centerline{\epsfbox{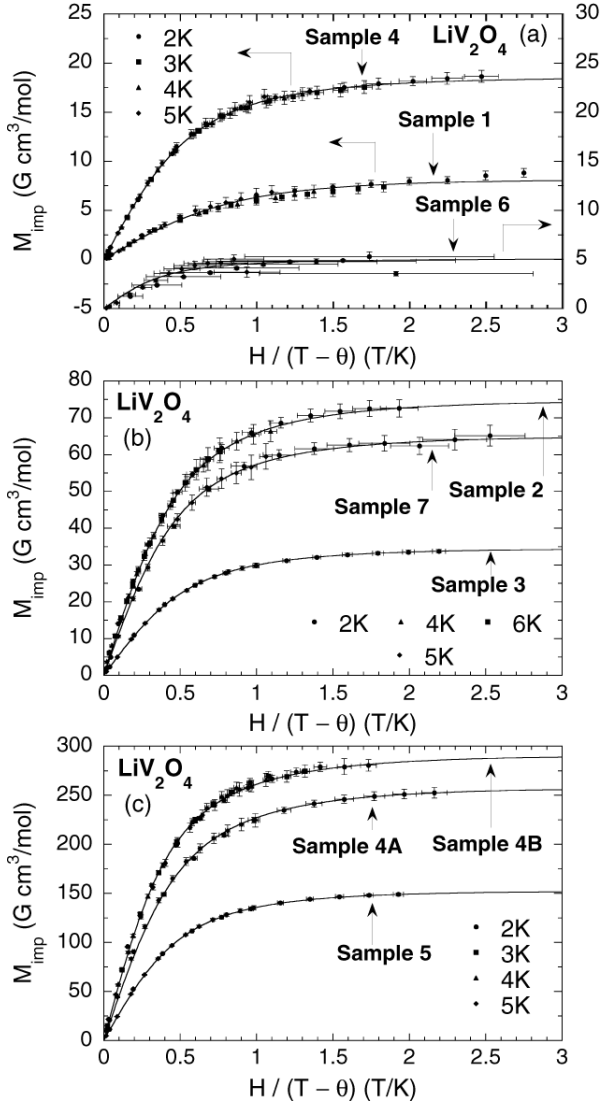}}
\vglue 0.1in
\caption{\mbox{Calculated impurity magnetizations} $M_{\rm imp} \equiv$ $M^{\rm
obs}-\chi H$ versus
$H/(T-\theta_{\rm imp})$ for the nine LiV$_2$O$_4$ samples.  For each sample, the
solid curve is the  best-fit Brillouin function Eq.~(\protect\ref{MEq}).}
\label{MimpFigs}
\end{figure}

The magnetic parameters for each sample, obtained as described above, are listed in
Table~\ref{IsothermFitTable}.  Plots of $M_{\rm imp}$ versus $H/(T-\theta_{\rm
imp})$ for the nine samples are given in Figs.~\ref{MimpFigs}(a), (b) and (c),
where an excellent universal behavior for each sample at different temperatures is
seen.  The two magnetically purest samples 1 and 6 have the largest relative
deviations of the data from the respective fit curves, especially at the larger
values of $H/(T-\theta_{\rm imp})$.
\noindent
\widetext
\begin{table}
\vglue -0.1in
\caption{Results of magnetization $M^{\rm obs}(H,T)$ isotherm analyses, where the
$T$ values used are listed in the second column.  $f_{\rm mag~imp}$ is the molar
magnetic impurity concentration.  The error in the last digit of a quantity is
given in parentheses.  All numbers without an error listed were fixed in the fit. 
The Curie constant of the impurities was calculated from $C_{\rm imp}=f_{\rm
mag~imp}N_{\rm A}g^{2}_{\rm imp}\mu^{2}_{\rm B}S_{\rm imp}(S_{\rm imp}+1)/(3k_{\rm
B})$.}
\label{IsothermFitTable}
\begin{tabular}{ccccccccc} Sample & $T$ & $S_{\rm imp}$ & $g_{\rm imp}$ &
$\theta_{\rm imp}$ & $f_{\rm mag~imp}$ & $C_{\rm imp}$ & $\chi(2\,{\rm K})$ &
$d\chi/dT$\\ No. & (K) & (fixed) & & (K) & (mol\,\%) & ($10^{-3}$\,$\frac{{\rm
cm}^3\,{\rm K}}{{\rm mol}}$) & ($10^{-2}$\,$\frac{{\rm cm}^3}{{\rm mol}}$) &
($\frac{{\rm cm}^3}{{\rm mol\,K}}$)\\ \hline 1 & 2,3,4,5 & 3/2 & 2 & 0 & 0.049(2)
& 0.74 & 1.026(1) & 7.3(1)\\ 2 & 2,4,6 & 3 & 2.00(6) & $-$0.6(2) & 0.22(1) & 13 &
1.034(5) & 6.7(4)\\ 3 & 2,5 & 5/2 & 2.10(2) & $-$0.51(5) & 0.118(2) & 4.9 &
0.9979(6) & 7.46(7)\\ 4 & 2,3,4,5 & 5/2 & 2 & $-$0.2(1) & 0.066(2) & 2.5 &
0.9909(9) & 6.7(1)\\ 4A & 2,5 & 3 & 2 & $-$0.5(1) & 0.77(2) & 46 & 1.145(9) &
6.5(9)\\ 4B & 2,3,4,5 & 7/2 & 2 & $-$1.2(1) & 0.74(2) & 52 & 1.13(1) & 4.4(7)\\ 5
& 2,5 & 5/2 & 2.31(3) &
$-$0.59(4) & 0.472(8) & 24 & 1.091(2) & 5(3)\\ 6 & 2,5 & 4 & 2 & $-$0.9(14) &
0.0113(6) & 1.1 & 1.067 & 5.6(2)\\ 7 & 2,5 & 3 & 2 & $-$0.2(2) & 0.194(7) & 12 &
1.094(4) & 5.4(4)\\
\end{tabular}
\end{table}
\narrowtext\noindent
  Since these two samples contain extremely
small amounts of paramagnetic saturable impurities, the magnetic parameters of the
impurities could not be determined to high precision.  The impurity spins $S_{\rm
imp}$ obtained for the nine samples vary from 3/2 to 4.  In general, the magnetic
impurity Weiss temperature $|\theta_{\rm imp}|$ increased with magnetic impurity
concentration $f_{\rm imp}$.  From the chemi- cal analyses of the starting materials
(V$_2$O$_5$, NH$_4$VO$_3$ and Li$_2$CO$_3$) supplied by the manufacturer, magnetic
impurity concentrations of 0.0024\,mol\,\%\,Cr and 0.0033\,mol\,\%\,Fe are
inferred with respect to a mole of LiV$_2$O$_4$, which are too small to account
for the paramagnetic impurity concentrations we derived for our samples.

\subsubsection{Magnetization versus Temperature
Measurements}\label{SusceptibilitySec}

\begin{figure}
\epsfxsize=3.3in
\centerline{\epsfbox{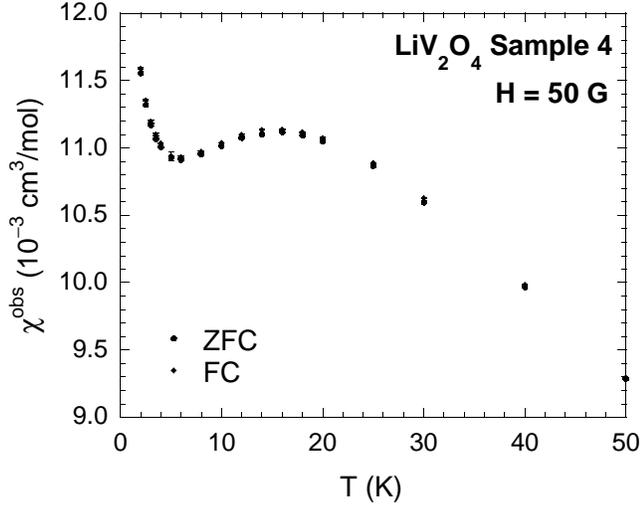}}
\vglue 0.1in
\caption{\mbox{Observed magnetic susceptibility} $\chi^{\rm obs}(T) \equiv$ 
$M^{\rm obs}(T)/H$ versus temperature $T$ in a low magnetic field $H=50$\,G of
LiV$_2$O$_4$ sample 4 cooled in zero field (ZFC) and in the low field (FC).}
\label{ZfcfclowHSample4}
\end{figure}

\paragraph{Low Magnetic Field ZFC and FC Measurements}\label{LowHChiSec} The
zero-field-cooled (ZFC)
$\chi^{\rm obs}(T)$ data in Fig.~\ref{ObsChiFigs}(a) for our highest magnetic
purity samples 1 and 6 show a broad maximum at $T^{\rm peak} \approx 16$\,K\@. 
One interpretation might be that static short-range (spin-glass) ordering sets in
below this temperature.  To check for spin-glass ordering, we carried out
low-field (10--100\,G) ZFC and field-cooled (FC) magnetization measurements from
1.8--2\,K to 50\,K on all samples except samples 2 and 4B.  For each sample, there
was no hysteresis between the ZFC and FC measurements, as illustrated for sample 4
in Fig.~\ref{ZfcfclowHSample4}, and thus no evidence for spin-glass ordering above
1.8--2\,K\@.\cite{Mydosh1993}

Ueda {\em{et al}}.\cite{Ueda1997} reported that spin-glass ordering occurs in the
zinc-doped lithium vanadium oxide spinel Li$_{1-x}$Zn$_x$V$_2$O$_4$ for $0.1 < x
\leq 0.9$\,. However, spin-glass ordering was not seen in the pure compound
LiV$_2$O$_4$, consistent with our results.  Further, positive-muon spin relaxation
$\mu$SR measurements for sample 1 did not detect static magnetic ordering down to
20\,mK\@.\cite{Kondo1997}  However, the $\mu$SR measurements did indicate the
presence of static spin-glass ordering in the off-stoichiometric sample 3 below
0.8\,K\@.\cite{Kondo1997}  As mentioned in Sec.~\ref{StructureSec}, the
stoichiometry of sample 3 was intentionally made slightly cation-
\newpage
\begin{figure}
\epsfxsize=3.3in
\centerline{\epsfbox{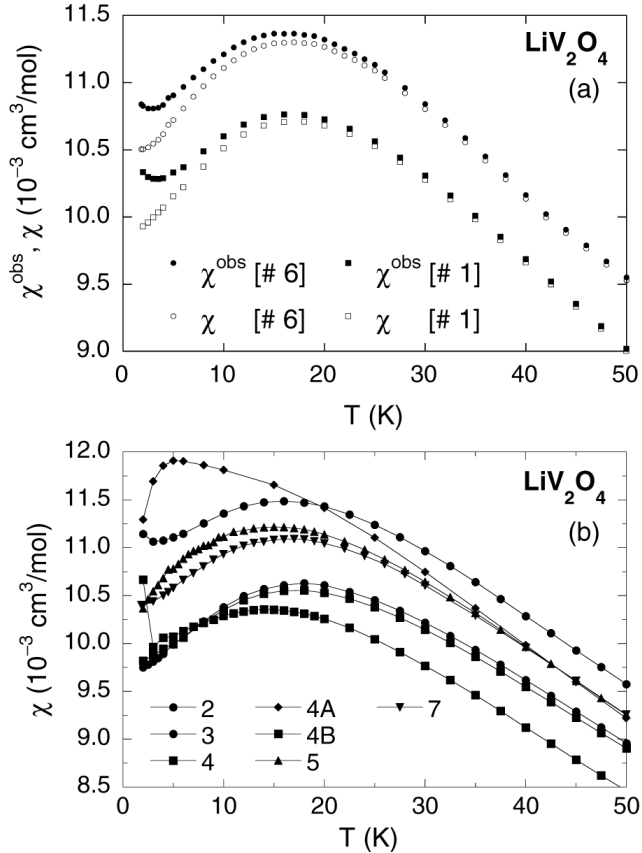}}
\vglue 0.1in
\caption{Observed susceptibilities $\chi^{\rm obs}$ and derived intrinsic
susceptibilities $\chi$ versus temperature $T$ of (a) samples 1 and 6 and (b)
samples 2, 3, 4, 5, 7, 4A and 4B.  The solid lines are guides to the eye.}
\label{ChiAllFigs}
\end{figure}
\noindent
deficient, and may contain cation vacancies.  Such a defective structure could
facilitate the occurrence of the spin-glass behavior by relieving the geometric
frustration among the V spins.  Whether the nature of the spin-glass ordering in
sample 3 is similar to or different from that in Li$_{1-x}$Zn$_{x}$V$_2$O$_4$
noted above is at present unclear.
\paragraph{Intrinsic Susceptibility}\label{IntChiSec}

The intrinsic susceptibility $\chi(T)$ was derived from the observed $M^{\rm
obs}(T)$ data at fixed $H=$ 1\,T using $\chi(T) = [M^{\rm obs}(T) - M_{\rm
imp}(H,T)]/H$, where $M_{\rm imp}(H, T)$ is given by Eq.~(\ref{MEq}) with
$H=1$\,T and by the parameters for each sample given
in Table~\ref{IsothermFitTable}, and $T$ is the only  variable.  The
$\chi(T)$ for each of the nine samples is shown in Figs.~\ref{ChiAllFigs}(a) and
(b), along with $\chi^{\rm obs}(T)$ for samples 1 and 6.  A shallow broad peak in
$\chi(T)$ is seen at a temperature $T_{\rm peak}=18, 16, 18, 18, 15, 17, 17, 5$
and 14\,K for samples 1--7, 4A and 4B, respectively.  The peak profiles seen in
$\chi(T)$ for the two magnetically purest samples 1 and 6 are regarded as most
closely reflecting the intrinsic susceptibility of LiV$_2$O$_4$.  This peak shape
is obtained in the derived $\chi(T)$ of all the samples except for sample 4A, as
seen in Fig.~\ref{ChiAllFigs}(b).  The physical nature of the magnetic impurities
in sample 4A is evidently different from that in the other samples.  Except for
the anomalous sample 4A, the $\chi(T=0)$ values were estimated from
Figs.~\ref{ChiAllFigs}(a) and (b), neglecting the small residual increases at the
lowest $T$ for samples 2, 6, 7 and 4B, to be
\begin{eqnarray}
\chi(0)=&9.8&,~10.8,~9.6,~9.7,~10.0,~10.2,~10.2,~\nonumber\\
&9.8& \times 10^{-3}\,{\rm cm^3/mol}~~(\mbox{samples~1--7,~4B})~~.
\label{ChiZeroEq}
\end{eqnarray}

\section{Modeling of the Intrinsic Magnetic Susceptibility}\label{ModelSec}

\subsection{The Van Vleck Susceptibility}\label{VVChiSec} The Van Vleck
paramagnetic orbital susceptibility $\chi^{\rm VV}$ may be obtained in favorable
cases from the so-called $K$-$\chi$ analysis, {\em i.e.}, if the transition metal
NMR frequency shift $K$ depends linearly on $\chi$, with
$T$ an implicit parameter.  One decomposes $\chi(T)$ per mole of transition metal
atoms according to
$\chi(T) = \chi^{\rm core} + \chi^{\rm VV} + \chi^{\rm spin}(T)$.  We neglect the
diamagnetic orbital Landau susceptibility, which should be small for $d$-electron
bands.\cite{White1983}  The NMR shift is written in an analogous fashion as
\begin{equation} K(T) = K^{\rm VV} + K^{\rm spin}(T)~~;
\label{EqKSum}
\end{equation} a term $K^{\rm core}$ does not appear on the right-hand side of
Eq.~(\ref{EqKSum}) because the absolute shift due to $\chi^{\rm core}$ is expected
to be about the same as in the Knight shift reference compound and hence does not
appear in the shift measured with respect to the reference compound.  Each
component of $K$ is written as a product of the corresponding component of $\chi$
and of the hyperfine coupling constant $A$ as
\begin{mathletters}
\label{EqK2:all}
\begin{eqnarray}
K^{\rm VV} &=& \frac{A^{\rm VV}}{N_{\rm A}\mu_{\rm B}}\chi^{\rm VV}~~,\\
\label{EqK2:a}
K^{\rm spin} &=& \frac{A^{\rm spin}}{N_{\rm A}\mu_{\rm B}}\chi^{\rm spin}~~.
\label{EqK2:b}
\end{eqnarray}
\end{mathletters}
Combining Eqs.~(\ref{EqKSum}) and (\ref{EqK2:all}) yields
\begin{equation} K = \frac{A^{\rm VV}}{N_{\rm A}\mu_{\rm B}}\chi^{\rm VV} +
\frac{A^{\rm spin}}{N_{\rm A}\mu_{\rm B}}\chi^{\rm spin}~~.
\label{EqK3}
\end{equation} If $K(T)$ varies linearly with $\chi(T)$, then the slope is $A^{\rm
spin}/N_{\rm A}\mu_{\rm B}$ since $\chi^{\rm VV}$ (and $\chi^{\rm core}$) is
normally independent of $T$.  We write the observed linear relation as
\begin{equation} K = K_o + \frac{A^{\rm spin}}{N_{\rm A}\mu_{\rm B}}\chi~~.
\label{EqKChi}
\end{equation} Setting the right-hand-sides of Eqs.~(\ref{EqK3}) and
(\ref{EqKChi}) equal to each other gives
\begin{equation}
\chi^{\rm VV} = \frac{N_{\rm A}\mu_{\rm B}K_o + A^{\rm spin}\chi^{\rm
core}}{A^{\rm VV} - A^{\rm spin}}~~.
\label{EqChiVV}
\end{equation}
\begin{figure}
\epsfxsize=3.1in
\centerline{\epsfbox{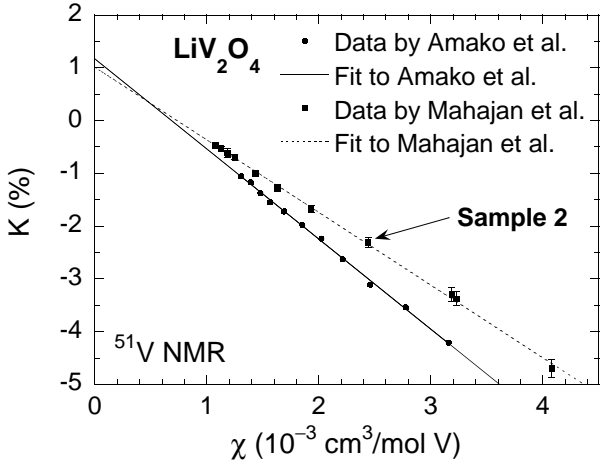}}
\vglue 0.1in
\caption{$^{51}$V NMR Knight shift $K$ versus observed magnetic susceptibility
$\chi^{\rm obs}$ for LiV$_2$O$_4$ by Amako {\em et
al.}\protect\cite{Takagi1987,Amako1990} and by Mahajan {\em et
al.}\protect\cite{Mahajan1998} for LiV$_2$O$_4$ sample 2.  The lines are linear
fits to the data according to Eq.~(\protect\ref{EqKChiFit}).}
\label{FigKChi}
\end{figure}

From $^{51}$V NMR and $\chi(T)$ measurements, the $K$ {\em vs.}\ $\chi$
relationship for LiV$_2$O$_4$ was determined by Amako {\em et al.}\cite{Amako1990}
and was found to be linear from 100--300\,K, as shown in Fig.~\ref{FigKChi}.  Our
fit to their data gave
\begin{equation} K = 0.0117(4) - \biggl[17.08(21)\,{\rm
\frac{mol\,V}{cm^3}}\biggr]\chi\biggl({\rm
\frac{cm^3}{mol\,V}}\biggr)~~,
\label{EqKChiFit}
\end{equation} shown as the straight line in Fig.~\ref{FigKChi}.  Comparison of
Eqs.~(\ref{EqKChi}) and (\ref{EqKChiFit}) yields
\begin{mathletters}
\label{EqLiK:all}
\begin{equation}
K_o = 0.0117(4)~~,
\label{EqLiK:a}
\end{equation}
\begin{equation}
A^{\rm spin} = -95.4(12)\,{\rm kG}~~.
\label{EqLiK:b}
\end{equation}
\end{mathletters}

The orbital Van Vleck hyperfine coupling constants for V$^{+3}$ and V$^{+4}$ are
similar.  For atomic V$^{+3}$, one has $A^{\rm VV}=403$\,kG
(Ref.~\onlinecite{Takigawa1996}).  We will assume that $A^{\rm VV}$ in
LiV$_2$O$_4$ is given by that\cite{Pouget1972} for atomic V$^{+4}$,
\begin{equation} A^{\rm VV} = 455\,{\rm kG}~~.
\label{EqAVV}
\end{equation} The core susceptibility is estimated here from Selwood's
table,\cite{Selwood1956} using the contributions [in units of 
$-10^{-6}$\,cm$^3$/(mol~ion)] 1 for Li$^{+1}$, 7 for V$^{+4}$ and 12 for O$^{-2}$,
to be
\begin{equation}
\chi^{\rm core} = -63 \times 10^{-6}\,\frac{{\rm cm^3}}{{\rm mol}}~~.
\label{EqChiCore}
\end{equation}
Inserting Eqs.~(\ref{EqLiK:all})--(\ref{EqChiCore}) into (\ref{EqChiVV}) yields
\begin{equation}
\chi^{\rm VV} = 2.48(9) \times 10^{-4}\,\frac{{\rm cm^3}}{{\rm mol}}~~.
\label{EqChiVV2}
\end{equation}

Mahajan {\em et al.}\cite{Mahajan1998} have measured the $^{51}$V $K(T)$ for our
LiV$_2$O$_4$ sample 2 from 78 to 575\,K\@.  Their data are plotted versus our
measurement of $\chi^{\rm obs}(T)$ for sample 2 from 74 to 400\,K in
Fig.~\ref{FigKChi}.  Applying the same $K$-$\chi$ analysis as above, we obtain
\begin{equation} K_o=0.0101(3)~~,
\end{equation}
\begin{equation} A^{\rm spin}=-76.9(8)\,{\rm kG}~~,
\end{equation}
\begin{equation}
\chi^{\rm VV}=2.22(6)\times10^{-4}\,\frac{{\rm cm^3}}{{\rm mol}}~~,
\label{ChiVVMahajanEq}
\end{equation} where the linear fit of $K$ {\em vs.} $\chi^{\rm obs}$ is shown by
the dashed line in Fig.~\ref{FigKChi}.

We may compare our similar values of $\chi^{\rm VV}$ for LiV$_2$O$_4$ in
Eqs.~(\ref{EqChiVV2}) and (\ref{ChiVVMahajanEq}) with those obtained from
$K$-$\chi$ analyses of other oxides containing V$^{+3}$ and V$^{+4}$.  For
stoichiometric V$_2$O$_3$ above its metal-insulator transition temperature of
$\sim 160$\,K, Jones\cite{Jones1965} and Takigawa {\em et~al.}\cite{Takigawa1996}
respectively obtained
$\chi^{\rm VV} = 2.10$ and $2.01 \times 10^{-4}$\,cm$^3$/(mol~V).  Kikuchi {\em et
al.}\cite{Kikuchi1996} obtained
$\chi^{\rm VV} = 0.92 \times 10^{-4}$\,cm$^3$/(mol~V) for LaVO$_3$, and for
VO$_2$, Pouget {\em et al.}\cite{Pouget1972} obtained $\chi^{\rm VV} = 0.65 \times
10^{-4}$\,cm$^3$/(mol~V).  

\subsection{High-Temperature Series Expansion Analysis of the
Susceptibility}\label{HTSESec}

\begin{figure}
\epsfxsize=3.1in
\centerline{\epsfbox{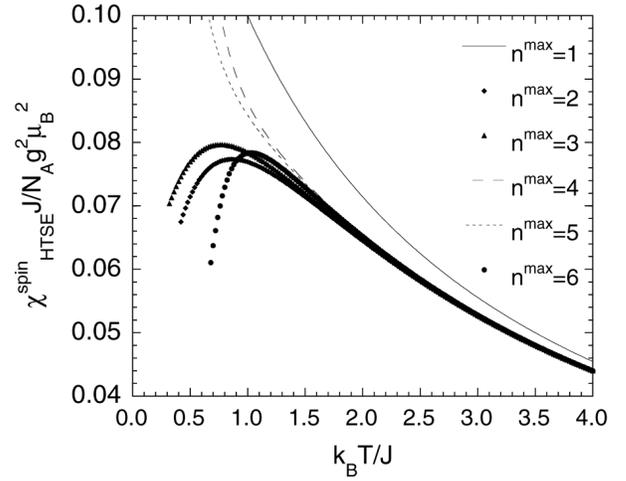}}
\vglue 0.1in
\caption{High-temperature series expansion predictions of the normalized spin
susceptibility $\chi^{\rm spin}_{\rm HTSE}J/N_{\rm A}g^2\mu_{\rm B}^2$ with
$n^{\rm max}=1\mbox{--}6$ versus reduced temperature
$k_{\rm B}T/J$ [Eq.~(\protect\ref{HTSEChiSpinEq})] for the antiferromagnetically
coupled spins $S=1/2$ in the $B$ sublattice of a normal-spinel compound
$A[B_2]$O$_4$.}
\label{HTSEPlot}
\end{figure}

Above $\sim 50$\,K the monotonically decreasing susceptibility of LiV$_2$O$_4$
with increasing $T$ has been interpreted by previous workers in terms of the
Curie-Weiss law for a system of spins $S=1/2$ and $g\approx2$.
\cite{Kessler1971,Nakajima1991,ChamberlandHewston1986,Takagi1987,%
Hayakawa1989,Johnston1995} 
To extend this line of analysis, we have fitted $\chi(T)$ by the high-temperature
series expansion (HTSE) prediction\cite{Rushbrooke1958,HTSE1974} up to sixth order
in $1/T$.  The assumed nearest-neighbor (NN) Heisenberg Hamiltonian between
localized moments reads ${\cal H}=J\sum_{\langle i,j \rangle}S_i \cdot S_j$, where
the sum is over all NN pairs, $J$ is the NN exchange coupling constant and $J>0$
denotes AF interactions. A HTSE of $\chi^{\rm spin}_{\rm HTSE}(T)$ arising from
this Hamiltonian up to the $n^{\rm max}$-th order of $J/k_{\rm B}T$ for general
lattices and spin $S$ was determined by Rushbrooke and Wood,\cite{Rushbrooke1958}
given per mole of spins by
\begin{equation}
\frac{N_{\rm A}g^2\mu_{\rm B}^2}{\chi^{\rm spin}_{\rm HTSE}(T)J}={\frac{3k_{\rm
B}T}{S(S+1)J}}\sum_{n=0}^{n^{\rm max}} b_n\biggl(\frac{J}{k_{\rm B}T}\biggr)^n~~,
\label{HTSEChiSpinEq}
\end{equation}
where $b_0\equiv 1$.  The $b_n$ coefficients for $S=1/2$ up to sixth order
($n^{\rm max}=6$) are
\begin{eqnarray}
b_1=\frac{z}{4}~,~~~b_2=\frac{z}{8}&,&~~~b_3=\frac{z}{24}(1-\frac{5p_1}{8})~~,
\nonumber\\
b_4=\frac{z}{768}(13-5z&-&15p_1+5p_2)~~,\nonumber\\
b_5=-\frac{z}{15360}(90z&-&122+245p_1-60zp_1-45p_1^2\nonumber\\
&-&90p_2+25p_3)~~,\nonumber\\
b_6=\frac{z}{184320}(134z^2&-&783z+713+908zp_1-2697p_1\nonumber\\
&-&106zp_1^2+1284p_1^2-234zp_2\nonumber\\
&+& 849p_2-291p_3+75p_4-288p_1p_2\nonumber\\
&-&51q-8r)~~.
\label{HTSEcoeff}
\end{eqnarray}
Here $z$ is the nearest-neighbor coordination number, and $p_n$, $q$ and $r$ are
so-called lattice parameters which depend upon the geometry of the magnetic
lattice.  The Curie Law corresponds to maximum order $n^{\rm max}=0$, and the
Curie-Weiss Law to maximum order $n^{\rm max}=1$.  For the $B$ sublattice of a
normal-spinel structure compound $A[B_2]$O$_4$, which is geometrically frustrated
for AF interactions, the parameters are $z=6$, $p_1=2$, $p_2=2$, $p_3=0$,
$p_4=2$, $q=0$, $r=2$.  For $S=1/2$, Eq.~(\ref{HTSEcoeff}) then yields
\begin{eqnarray}
b_1&=&\frac{3}{2}~,~~~b_2=\frac{3}{4}~,~~~b_3=-\frac{1}{16}~,\nonumber\\
b_4&=&-\frac{37}{128}~,~~~b_5=\frac{43}{640}~,~~~b_6=\frac{1361}{6144}~~.
\label{HTSEcoeff2}
\end{eqnarray}
Figure~\ref{HTSEPlot} illustrates the HTSE predictions of
Eq.~(\ref{HTSEChiSpinEq}) for $S=1/2$ using these $b_n$ coefficients for $n^{\rm
max}=1$ to 6.  The theoretical $\chi^{\rm spin}_{\rm HTSE}(T)$ predictions with
$n^{\rm max}=2$, 3 and~6 exhibit broad maxima as seen in our experimental
$\chi(T)$ data.  The prediction with $n^{\rm max}=6$ is evidently accurate at
least for $k_{\rm B}T/J\gtrsim1.6$; at lower $T$, the theoretical curves with
$n^{\rm max}=5$ and 6 diverge noticably from each other on the scale of
Fig.~\ref{HTSEPlot}.  Our fits given below of the experimental data by the
theoretical $\chi^{\rm spin}_{\rm HTSE}(T)$ prediction were therefore carried out
over temperature ranges for which $k_{\rm B}T/J\gtrsim 1.6$.  The Weiss
temperature $\theta$ in the Curie-Weiss law is given for coordination number $z=6$
and $S=1/2$ by $\theta=-zJS(S+1)/(3k_{\rm B})=-3J/(2k_{\rm B})$.

To fit the HTSE calculations of $\chi^{\rm spin}_{\rm HTSE}(T)$ to experimental
data, we assume that the experimentally determined intrinsic susceptibility
$\chi(T)$ is the sum of a $T$-independent term
$\chi_0$ and $\chi^{\rm spin}_{\rm HTSE}(T)$,
\begin{figure}
\epsfxsize=3.1in
\centerline{\epsfbox{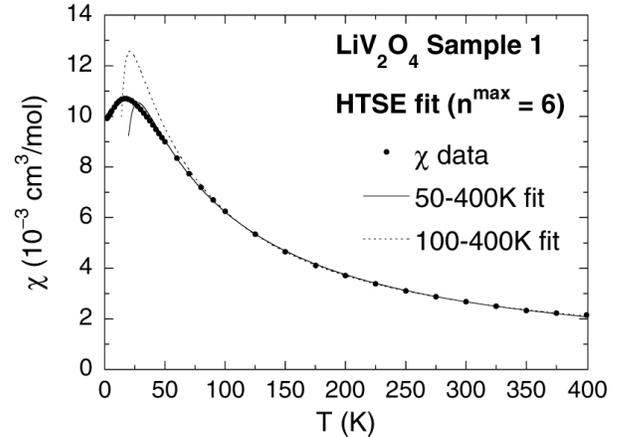}}
\vglue 0.1in
\caption{Intrinsic susceptibility $\chi$ versus temperature $T$ for LiV$_2$O$_4$
sample~1 (filled circles) and fits (curves) by the high-$T$ series expansion
(HTSE) prediction to 6th order in $1/T$ for the 50--400 and 100--400\,K
temperature ranges.}
\label{ChiHiTSample1}
\end{figure}
\noindent
\begin{equation}
\chi(T)=\chi_0+\chi^{\rm spin}_{\rm HTSE}(T)~~,
\label{EqHTSE}
\end{equation}
with $\chi^{\rm spin}_{\rm HTSE}(T)$ given by Eq.~(\ref{HTSEChiSpinEq}) and the
$b_{n}$ coefficients for $S=1/2$ in Eq.~(\ref{HTSEcoeff2}).  The three parameters
to be determined are $\chi_0$, $g$ and $J/k_{\rm B}$.  The fitting parameters for
samples 1--7, 4A and 4B using $n^{\rm max}=6$, and for sample 1 also using
$n^{\rm max}=2$ and 3, are given in Table~\ref{HiT-Results} for the 50--400 and
100-400\,K fitting ranges.  The fits for these two fitting ranges for sample 1 and
$n^{\rm max}=6$ are  shown in Fig.~\ref{ChiHiTSample1}.  Both $g$ and $J/k_{\rm
B}$ tend to decrease as the lower limit of the fitting range increases.  The HTSE
fits for all the samples yielded the ranges $C=N_{\rm A}g^{2}\mu_{\rm
B}^{2}/(4k_{\rm B})=0.36\mbox{--}0.48$\,cm$^3$\,K/(mol\,V) and $\theta=-20$ to
$-42$\,K, in agreement with those reported previously (see
Table~\ref{PastWorks}). 
$\chi_0$ was found to be sensitive to the choice of fitting temperature range. 
For the 50--400\,K range, $\chi_0$ was negative for some samples.  Recalling the
small negative value of the core diamagnetic contribution in
Eq.~(\ref{EqChiCore}) and the larger positive value of the Van Vleck
susceptibility in Eqs.~(\ref{EqChiVV2}) and (\ref{ChiVVMahajanEq}), it is
unlikely that
$\chi_0$ [defined below in Eq.~(\ref{ChiZeroEq2})] would be negative.  Negative
values of $\chi_0$ occur when the low-$T$ limit of the fitting range is below
100\,K, and may therefore be an artifact of the crossover between the local moment
behavior at high 
$T$ and the HF behavior at low $T$.

\begin{figure}
\epsfxsize=3.1in
\centerline{\epsfbox{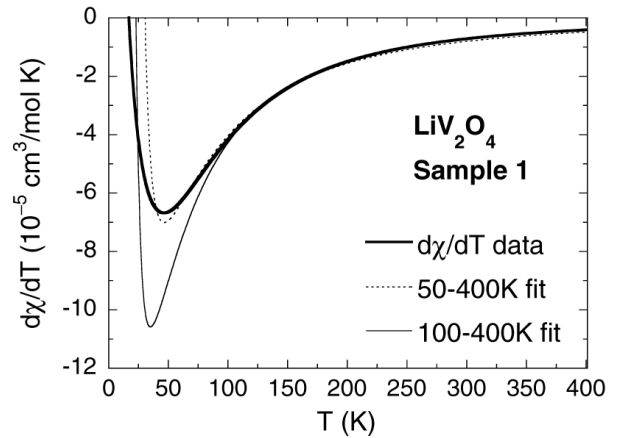}}
\vglue 0.1in
\caption{Temperature derivative of the experimental intrinsic susceptibility,
$d\chi/dT$, for LiV$_2$O$_4$ sample 1 (heavy solid curve).  Fits by the $T$
derivative of the HTSE prediction $d\chi^{\rm spin}/dT$ in
Eq.~(\protect\ref{HTSEChiSpinEq}) are also shown for $T$ ranges of  50--400\,K
(dashed curve) and 100--400\,K (light solid curve).}
\label{fitderivHTSE}
\end{figure}

\begin{figure}
\epsfxsize=3.3in
\centerline{\epsfbox{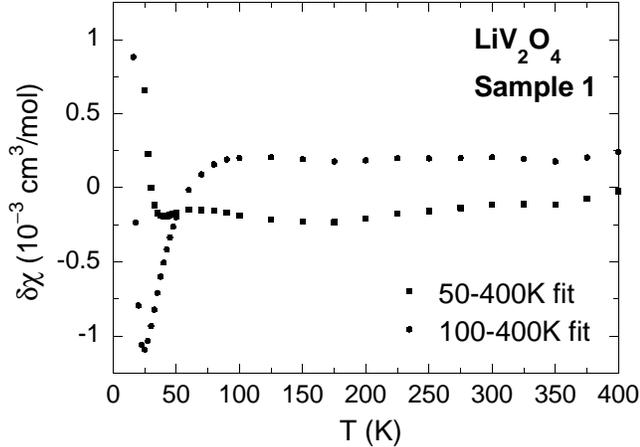}}
\vglue 0.1in
\caption{The differences between the experimental intrinsic susceptibility
$\chi(T)$ of LiV$_2$O$_4$ sample~1 and the HTSE prediction $\chi^{\rm spin}$
obtained from the $T$ derivative analysis,
$\delta\chi(T) \equiv\chi(T)-\chi^{\rm spin}$, versus temperature $T$ for the
fitting $T$ ranges of 50--400\,K (open squares) and 100--400\,K (filled circles). 
For a valid fit, these differences should be the $T$-independent susceptibility
$\chi_0$. }
\label{fitdiffChi}
\end{figure}

To eliminate $\chi_0$ as a fitting parameter, we also fitted ${\rm d}\chi/{\rm
d}T$ by the HTSE prediction for that quantity.  The experimental
${\rm d}\chi/{\rm d}T$ was determined from a Pad\'e approximant fit to $\chi(T)$
and is plotted in Fig.~\ref{fitderivHTSE} for sample~1.  These data were fitted
by ${\rm d}\chi^{\rm spin}_{\rm HTSE}/{\rm d}T$ obtained from the HTSE prediction
Eq.~(\ref{HTSEChiSpinEq}) with $n^{\rm max}= 6$, where the fitting parameters are
now $g$ and $J/k_{\rm B}$.  The fits were carried out over the same two $T$
ranges as in Fig.~\ref{ChiHiTSample1}; Table~\ref{derivHTSEResults} displays the
fitting parameters and the fits are plotted in Fig.~\ref{fitderivHTSE}.  Both $g$
and $J/k_{\rm B}$ were found to be larger than the corresponding values in
Table~\ref{HiT-Results}.  Of the two fitting ranges, the 100--400\,K fit is the
best fit inside the respective range, though it shows a large deviation from the
data below this range.  Using the fitting parameters, the HTSE $\chi^{\rm
spin}(T)$ is obtained from Eq.~(\ref{HTSEChiSpinEq}).  According to
Eq.~(\ref{EqHTSE}), the difference between the experimental $\chi(T)$ and
$\chi^{\rm spin}_{\rm HTSE}(T)$, $\delta\chi(T)=\chi(T)-\chi^{\rm spin}_{\rm
HTSE}(T)$, 
\widetext
\begin{table}
\caption{Results of high-temperature series expansion calculation fits to the
intrinsic magnetic susceptibility data for LiV$_2$O$_4$ over the temperature
ranges 50--400\,K and 100--400\,K\@.  The error in the last digit of a quantity is
given in parentheses.}
\label{HiT-Results}
\begin{tabular}{cccccccc}
Sample& $n^{\rm max}$ & \multicolumn{3}{c}{50--400\,K} &
\multicolumn{3}{c}{100--400\,K} \\ \cline{3-8} No. & & $\chi_0$ & $g$ & $J/k_{\rm
B}$& $\chi_0$ & $g$ &
$J/k_{\rm B}$\\
 & & ($10^{-4}$\,cm$^3$/mol) & & (K) & ($10^{-4}$\,cm$^3$/mol) & & (K)\\\hline 1 &
2 &        0.8(4) & 2.17(1) & 25.8(5) & 2.7(3) & 2.07(2) & 20(1) \\ 1 & 3 &    
0.7(4) & 2.18(2) & 26.2(6) & 2.6(3) & 2.07(2) & 20(1) \\ 1 & 6 &     0.5(4) &
2.19(2) & 26.9(7) & 2.6(3) & 2.07(2) & 20(1) \\ 2 & 6 & 
$-$0.2(5) & 2.26(2) & 26.7(8) & 2.6(3) & 2.11(2) & 19(1) \\ 3 & 6 &  $-$1.3(5) &
2.23(2) & 27.8(7) & 1.4(3) & 2.08(2) & 20.5(8) \\ 4 & 6 &        1.1(6) & 2.16(3)
& 26.4(9) & 4.1(5) & 1.99(3) & 17(2) \\ 4A & 6 & $-$0.6(8) & 2.20(3) & 26(1) & 
2.3(2) & 2.05(1) & 18.1(6) \\ 4B & 6 & $-$0.7(5) & 2.12(2) & 26.2(8) & 1.8(5) &
1.97(3) & 18(2) \\ 5 & 6 &      1.2(7) & 2.17(3) & 25(1) & 4.9(7) & 1.95(4) & 13(2)
\\ 6 & 6 &     0.8(1) & 2.251(6) & 26.5(2) & 3.3(7) & 2.108(4) & 18.4(2) \\ 7 & 6
&     0.5(3) & 2.20(1) & 25.8(5) & 3.0(1) & 2.051(8) & 17.5(4) \\
\end{tabular}
\end{table}
%
\begin{table}
\vglue-0.05in
\caption{Parameters $g$ and $J/k_{\rm B}$ obtained by fitting the temperature $T$ 
derivative of the experimental intrinsic susceptibility data for LiV$_2$O$_4$
samples 1 and 6 by the $T$ derivative of the HTSE spin susceptibility
[Eq.~(\protect\ref{HTSEChiSpinEq})] with $n^{\rm max}=6$ for two different
temperature ranges of the fit.  The $T$-independent susceptibility $\chi_0$ was
determined by averaging $\delta\chi(T)$, see Fig.~\protect\ref{fitdiffChi}.  The
error in the last digit of a quantity is given in parentheses.}
\label{derivHTSEResults}
\begin{tabular}{ccccccc} & \multicolumn{3}{c}{50--400\,K} &
\multicolumn{3}{c}{100--400\,K} \\
\cline{2-7} Sample No. & $\chi_0$ & $g$  & $(J/k_{\rm B})$ & $\chi_0$ & $g$  &
$(J/k_{\rm B})$ \\ 
& ($10^{-4}$\,$\frac{{\rm cm}^3}{{\rm mol}}$) & & (K) & ($10^{-4}$\,$\frac{{\rm
cm}^3}{{\rm mol}}$) & & (K)\\
\hline 
1 & $-$1.5(1) & 2.275(3) & 29.61(7)   & 2.00(4) & 2.103(2) & 22.27(8) \\ 
6 & $-$2.73(5) & 2.402(4) & 31.61(9)  & 2.11(1) & 2.174(3) & 22.1(1)  \\
\end{tabular}
\end{table}
\narrowtext
\noindent
should represent the $T$-independent contribution $\chi_0$. $\delta\chi(T)$ is
plotted for sample~1 versus $T$ in Fig.~\ref{fitdiffChi} for the 50--400\,K and
100--400\,K fit ranges.  Again, the superiority of the 100--400\,K fitting range
to the other is evident, {\em i.e.}, $\chi_0$ is more nearly constant for this
fitting range.  $\chi_0$ for the 50--400\,K fit range is negative within the
range.  This sign is opposite to that obtained in the 
\newpage\noindent
first HTSE fitting results
in Table~\ref{HiT-Results}.  This inconsistency found in the
 fit using a low $T$ limit below 100\,K may again be due to changing physics in
the crossover regime, which would invalidate the parameters.  By averaging the
$\chi_0$  values for samples 1 and 6 in the given ranges, we obtained the
$T$-independent contribution $\chi_0$, as listed in Table~\ref{derivHTSEResults}.
In the itinerant plus localized moment model implicitly assumed in this section,
$\chi_0$ can be decomposed as
\begin{equation}
\chi_0=\chi^{\rm core}+\chi^{\rm VV}+\chi^{\rm Pauli}~~,
\label{ChiZeroEq2}
\end{equation}
where $\chi^{\rm Pauli}$ is the temperature-independent Pauli spin
susceptibility of the conduction electrons.  Using the results of $\chi^{\rm
core}$ [Eq.~(\ref{EqChiCore})], $\chi^{\rm VV}$ [Eq.~(\ref{ChiVVMahajanEq})] and
$\chi_0$ (100--400\,K range, Table~\ref{derivHTSEResults}), we find
\begin{mathletters}
\label{EqChiPauli:all}
\begin{equation}
\chi^{\rm Pauli}=0.41(10)\times10^{-4}\,{\rm cm^3/mol}~~({\rm sample~1})~~,
\label{EqChiPauli:a}
\end{equation}
\begin{equation}
\chi^{\rm Pauli}=0.52(7)\times10^{-4}\,{\rm cm^3/mol}~~({\rm sample~6})~~.
\label{EqChiPauli:b}
\end{equation}
\end{mathletters}
These $\chi^{\rm Pauli}$ values are approximately four times smaller than that
obtained for LiV$_2$O$_4$ by Mahajan {\em et al.}\cite{Mahajan1998}  They used
$\chi^{\rm obs}(T)$ in the $T$ range 100--800\,K, combining our $\chi^{\rm
obs}(T)$ data to 400\,K with those of Hayakawa {\em et al.}\cite{Hayakawa1989} to
800\,K\@.  By fitting these combined data by the expression $\chi^{\rm
obs}(T)=\chi_0+2C/(T-\theta)$, they obtained
$\chi_0=5.45\times10^{-4}$\,cm$^{3}$/mol.  As shown above and also discussed in
Ref.~\onlinecite{Mahajan1998}, the value of $\chi_0$ is sensitive to the fitting
temperature range. For LiTi$_2$O$_4$, $\chi^{\rm Pauli}$ $\sim 2 \times
10^{-4}$\,cm$^3$/mol (Refs.~\onlinecite{Johnston1976,Heintz1989}) between 20 and
300\,K, which is a few times larger than we find for LiV$_2$O$_4$ from the
100--400\,K range fits (Table~\ref{derivHTSEResults}).

\subsection{Crystal Field Model}\label{CFTSec}

The ground state of a free ion with one 3$d$ electron is $^{2}D_{3/2}$ and has
five-fold orbital degeneracy.  The point symmetry of a V atom in LiV$_2$O$_4$ is
trigonal.  If we consider the crystalline electric field (CEF) seen by a V atom
arising from only the six nearest-neighbor oxygen ions, the CEF due to a perfect
oxygen octahedron is cubic ($O_h$ symmetry), assuming here point charges for the
oxygen ions.  In this CEF the degeneracy of the five $d$ orbitals of the vanadium
atom is lifted and the orbitals are split by an energy ``$10Dq$'' into a lower
orbital $t_{2g}$ triplet and a higher orbital $e_g$ doublet.  However, in
LiV$_2$O$_4$ each V-centered oxygen octahedron is slightly distorted along one of
the $<$111$>$ directions [see Fig.~\ref{spinel}(b)], as discussed in
Sec.~\ref{StructureSec}.  This distortion lowers the local symmetry of the V atom
to $D_{3d}$ (trigonal) and causes a splitting of the $t_{2g}$ triplet into an
$A_{1g}$ singlet and an $E_{g}$ doublet.  It is not clear to us which of the
$E_{g}$ or $A_{1g}$ levels become the ground state, and how large the splitting
between the two levels is.  These questions cannot be answered readily without a
knowledge of the magnitudes of certain radial integrals,\cite{Gerloch1970} and
are not further discussed here.\cite{Kondo1998}  However, this trigonal splitting
is typically about an order of magnitude smaller than $10Dq$.\cite{Krupicka1982} 
In the following, we will examine the predictions for $\chi(T)$ of a $d^1$ or
$d^2$ ion in a cubic CEF and compare with our experimental data for LiV$_2$O$_4$.

\begin{figure}
\epsfxsize=3.1in
\centerline{\epsfbox{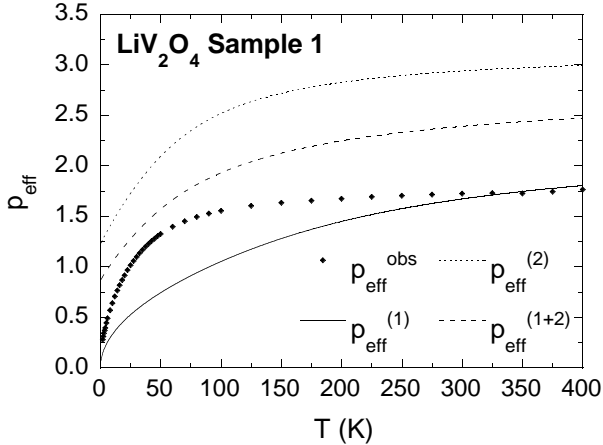}}
\vglue 0.1in
\caption{Observed effective magnetic moment in $\mu_{\rm B}$, $p_{\rm eff}^{\rm
obs}$, versus temperature $T$ of LiV$_2$O$_4$ sample 1 (filled diamonds). Also
shown as the curves are the predictions $p_{\rm eff}^{(1)}$ for $d^1$ ions and
$p_{\rm eff}^{(2)}$ for $d^2$ ions by Kotani,\protect\cite{Kotani1949} and $p_{\rm
eff}^{(1+2)}$ for an equal mixture of $d^1$ and $d^2$ ions, in a cubic crystalline
electric field, including spin-orbit coupling.}
\label{nfig}
\end{figure}

Kotani\cite{Kotani1949} calculated the effective magnetic moment $\mu_{\rm eff}
\equiv p_{\rm eff}\mu_{\rm B}$ per $d$-atom for a cubic CEF using the Van Vleck
formula.\cite{VanVleck1932}  The spin-orbit interaction is included, where the
coupling constant is $\lambda$.  For an isolated atom
$\mu_{\rm eff}(T)$ is defined by $\chi(T)\equiv N\mu_{\rm eff}^{2}(T)/(3k_{\rm
B}T)$, where $\mu_{\rm eff}$ is in general temperature-dependent and $N$ is the
number of magnetic atoms.  With spin included, one uses the double group for
proper representations of the atomic wavefunctions.  Then in this cubic double
group with one $d$-electron the six-fold (with spin) degenerate $t_{2g}$ level
splits into a quartet $\Gamma_{8}$($t_{2g}$) and a doublet
$\Gamma_{7}$($t_{2g}$).\cite{Kotani1949,Ballhausen1962,Dunn1965}  The four-fold
degenerate $e_{g}$ level does not split and its representation is
$\Gamma_{8}$($e_{g}$).  For a positive $\lambda$, as is appropriate for a $3d$
atom with a less than half-filled $d$-shell, $\Gamma_{8}$($t_{2g}$) is the ground
state, and the first-order Zeeman effect does not split it; this ground state is
non-magnetic.  Kotani does not include in his calculations of $\mu_{\rm eff}$ the
possible coupling of
$\Gamma_{8}$($t_{2g}$) and $\Gamma_{8}$($e_{g}$), which have the same symmetry,
and assumes that the cubic CEF splitting $10Dq$ is large enough to prevent
significant mixing.  On the other hand, the cubic double group with two
$d$-electrons gives an orbitally nondegenerate, five-fold spin-degenerate, ground
state with angular momentum quantum number ${\cal J}= 2$ which splits into five
non-degenerate levels under a magnetic field.  The spin-orbit coupling constant is
$\lambda=$ +250\,cm$^{-1}$ for $d^1$ (V$^{+4}$) and +105\,cm$^{- 1}$ for $d^2$
(V$^{+3}$).\cite{Dunn1961}  The effective moment is defined from the observed
molar susceptibility of LiV$_2$O$_4$ as $\chi^{\rm obs}(T)=\chi_0+2N_{\rm
A}[p_{\rm eff}^{\rm obs}(T)]^{2}\mu_{\rm B}^{2}/(3k_{\rm B}T)$, where we take
$\chi_0=2.00\times10^{-4}$\,cm$^3$/mol given in Table~\ref{derivHTSEResults}. 
Kotani's results from the Van Vleck equations are\cite{Kotani1949}
\begin{equation} p_{\rm
eff}^{(1)}=\Biggl[\frac{8+(3x-8)e^{-\frac{3}{2}x}}{x(2+e^{-\frac{3}{2}x})}
\Biggr]^{1/2}~~
\label{EqKotani1}
\end{equation} for the $d^1$ ion, and
\begin{equation} p_{\rm
eff}^{(2)}=\Biggr[\frac{3(\frac{5}{2}x+15+(\frac{x}{2}+9)e^{-x}-24
e^{-\frac{3}{2}x})}{x(5+3e^{-x}+e^{-\frac{3}{2}x})}\Biggr]^{1/2}~~
\label{EqKotani2}
\end{equation}
for the $d^2$ ion, where $x \equiv \lambda$/$k_{\rm B}T$.  Figure~\ref{nfig} shows
$p_{\rm eff}^{\rm obs}$, $p_{\rm eff}^{(1)}$ and $p_{\rm eff}^{(2)}$ as a function
of $T$.  For comparison is also shown $p_{\rm eff}^{(1+2)}$ obtained by assuming
that $p_{\rm eff}^{\rm obs}(T)$ arises from an equal mixture   of V$^{+3}$ and
V$^{+4}$ localized moments.  None of the three calculated curves agree with the
experimental data over the full temperature range.  However, in all three
calculations $p_{\rm eff}$ increases with $T$, in qualitative agreement with the
data, perhaps implying the importance of orbital degeneracy in LiV$_2$O$_4$ and/or
antiferromagnetic coupling between vanadium spins.  The nearly $T$-independent
$p_{\rm eff}^{\rm obs} \approx 1.8$ for $T\gtrsim 100$\,K is close to the
spin-only value $p_{\rm eff} = g\sqrt{S(S+1)}$ with $S= 1/2$ and $g \approx 2$, as
expected in the {\it absence} of orbital degeneracy; however, this result also
arises in the theory for the $d^1$ ion when $k_{\rm B}T\sim \lambda$, as seen by
comparison of the solid curve with the data in Fig.~\ref{nfig} at $\sim 300$\,K\@.

\subsection{Spin-1/2 Kondo Model and Coqblin-Schrieffer
Model}\label{KondoCSModelsSec}

$\chi^{\rm obs}(T)$ data for $f$-electron HF compounds are often found to be
similar to the predictions of the single-ion Kondo
model\cite{Wilson1975,Krishna1980,Krishna1980b,Rajan1982,Tsvelick1983} for spin
$S=$1/2 or its  extention to $S >$ 1/2 in the Coqblin-Shrieffer
model.\cite{Coqblin1969,Rajan1983} The zero-field impurity susceptibility
$\chi_{\rm CS}(T)$ of the Coqblin-Shrieffer model was calculated exactly as a
function of temperature by Rajan.\cite{Rajan1983}  His numerical results
$\chi_{\rm CS}(T)$ for impurity angular momentum quantum number ${\cal J} =
1/2$,~\ldots, 7/2 show a Curie-Weiss-like $1/T$ dependence (with logarithmic
corrections) for $T \gg T_{\rm K}$, where $T_{\rm K}$ is the Kondo temperature. 
As $T$ decreases, $\chi_{\rm CS}(T)$ starts to deviate from the $1/T$ dependence,
shows a peak (at $T\approx 0.2T_{\rm K}$) only for ${\cal J} \geq 3/2$, and levels
off for $T\lesssim 0.2T_{\rm K}$ for all ${\cal J}$.

In the zero temperature limit the molar susceptibility for ${\cal J}=S=1/2$ (which
corresponds to the $S=1/2$ Kondo model) is\cite{Rajan1983}
\begin{equation}
\chi_{\rm CS}(T = 0) = \frac{0.102678N_{\rm A}g^{2}\mu^{2}_{\rm
B}}{k_{\rm B}T_{\rm K}}~~.
\label{zeroTchiKondo}
\end{equation}
Setting $g=2$, and using the intrinsic $\chi(T\,\rightarrow\,0) =
0.0049$\,cm$^{3}$/(mol\,V) for LiV$_2$O$_4$ sample~1 from Eq.~(\ref{ChiZeroEq}),
Eq.~(\ref{zeroTchiKondo}) yields the Kondo temperature
\begin{equation}
T_{\rm K} = 32.1\,{\rm K}~~.
\label{TKondo1}
\end{equation}
On the other hand, if the $g$-value of 2.10 from Table~\ref{derivHTSEResults}
(100--400\,K range) is employed instead, the Kondo temperature is
\begin{equation}
T_{\rm K} = 35.5\,{\rm K}~~.
\label{TKondo2}
\end{equation}

\begin{figure}
\epsfxsize=3.1in
\centerline{\epsfbox{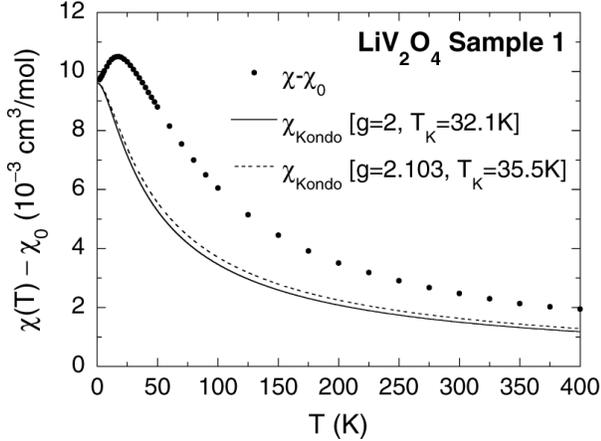}}
\vglue 0.1in
\caption{Temperature $T$-dependent part of the magnetic susceptibility,
$\chi-\chi_0$, versus $T$ for LiV$_2$O$_4$ sample 1 (filled circles).  Also shown
as solid and dashed curves are the predictions of the spin $S=1/2$ Kondo model for
$(g,\,T_{\rm K})=(2$,\,32.1\,K) and (2.103,\,35.5\,K), respectively, where $g$ is
the $g$-factor and $T_{\rm K}$ is the Kondo temperature.}
\label{JerezDataAndFitJ=1/2b}
\end{figure}

The temperature dependence of the impurity susceptibility of the $S=1/2$ Kondo
model was obtained using accurate Bethe ansatz calculations by Jerez and 
Andrei.\cite{JerezAndrei1997} Their $T\to 0$ value for the coefficient on the
right-hand-side of Eq.~(\ref{zeroTchiKondo}) is 0.1028164, about 0.1\% too
high compared with the correct prefactor in Eq.~(\ref{zeroTchiKondo}).  We fitted
their calculated values for $t = 0.00104$ to~102.53 by 
\begin{mathletters}
\label{EqKondoChiFit:all}
\begin{equation}
\frac{4\chi_{\rm CS} k_{\rm B}T}{Ng^{2}\mu^{2}_{\rm
B}}=\frac{1 +
{n_1\over t} + {n_2\over t^2} + {n_3\over t^3} + {4(0.1028164)n_5\over t^5}}{1 +
{d_1\over t} + {d_2\over t^2} + {d_3\over t^3} + {d_4\over t^4} + {n_5\over
t^6}}~~,
\label{EqKondoChiFit:a}
\end{equation}
\begin{eqnarray}
n_1 &=& 530.417~,~~~n_2 = 4697.91~,~~~n_3 = 1404.18~,\nonumber\\ 
n_5 &=& -418.781~,~~~d_1 = 695.557~,~~~d_2 = 8605.97~,\nonumber\\
d_3 &=& 11373.7~,~~~d_4 = 2937.88~,
\label{EqKondoChiFit:b}
\end{eqnarray}
\end{mathletters}
where $t \equiv T/T_{\rm K}$.  Equation~(\ref{EqKondoChiFit:a}) has the correct
form $\chi_{\rm CS}(0) + b t^2$ at low $T$ and approaches a Curie law in the
high-$T$ limit, as required by the Kondo model.  The large $n_i$ and $d_i$
coefficients arise because $\chi_{\rm CS}(T)$ converges very slowly to the Curie
law at high temperatures.   The rms deviation of the fit values from the Bethe
ansatz calculation values is 0.038\%, and the maximum deviation is 0.19\% at
$t=66.9$.  Using the above-stated $g$-values and $T_{\rm K}$ from
Eqs.~(\ref{TKondo1}) and (\ref{TKondo2}), the $S = 1/2$ $\chi_{\rm CS}(T)$
calculations are compared with our $\chi(T)$ data in
Fig.~\ref{JerezDataAndFitJ=1/2b}.  Note that in Fig.~\ref{JerezDataAndFitJ=1/2b},
both the $T$-independent $\chi_0$ (Table~\ref{derivHTSEResults}) and impurity
susceptibilities are already subtracted from $\chi^{\rm obs}$.  Although the
$T_{\rm K}$ values in Eqs.~(\ref{TKondo1}) and (\ref{TKondo2}) are comparable to
those obtained from specific heat analyses,\cite{Kondo1997,Johnston1998} the
$S=1/2$ Kondo model predictions for $\chi(T)$ with these $T_{\rm K}$ values do
not agree with our observed temperature dependence.  This failure is partly due
to the fact that our
$\chi(T)$ data exhibit a weak maximum whereas the $S=1/2$ Kondo model calculation
does not.

\begin{figure}
\epsfxsize=3.1in
\centerline{\epsfbox{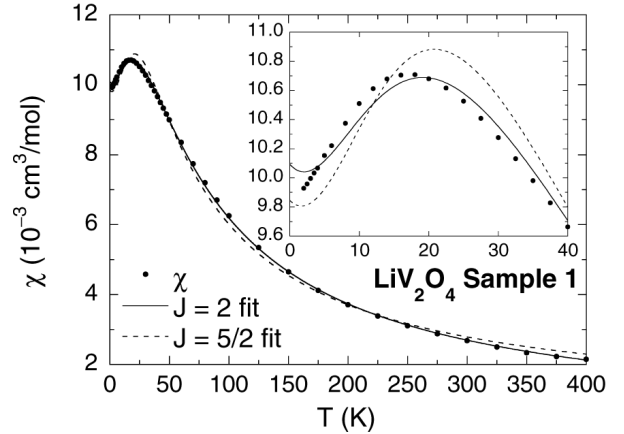}}
\vglue 0.1in
\caption{Intrinsic magnetic susceptibility $\chi$ of sample~1 versus temperature
$T$ and fits by the Coqblin-Schrieffer model prediction for spins ${\cal J}=2$
and 5/2.  The inset shows an expanded plot of the data and fits below 40\,K\@.}
\label{CoqblinFitFig}
\end{figure}

As noted above, the Coqblin-Schrieffer model for ${\cal J} \geq 3/2$ does give a
peak in $\chi_{\rm CS}(T)$.\cite{Rajan1983}   Defining the ratio
\begin{equation}
r (\%) = 100\,\frac{\chi^{\rm peak}_{\rm CS}- \chi_{\rm CS}(0)}{\chi_{\rm
CS}(0)}~~,
\end{equation}
where $\chi^{\rm peak}_{\rm CS}$ is the value of $\chi_{\rm CS}(T)$ at the peak,
the calculations\cite{Rajan1983} give $r=2$, 7, 11, 17 and 22\% for ${\cal
J}=3/2$, 2, 5/2, 3 and 7/2, respectively.  The observed value is $r=8.2$\% in
sample~1, which is between the theoretical values for ${\cal J} = 2$ and 5/2. 
Fits of $\chi_{\rm CS}(T)$ to our $\chi(T)$ data of sample~1 for $T=2$--400\,K are
shown in Fig.~\ref{CoqblinFitFig} and the parameters are
\begin{mathletters}
\label{CoqblinFitResults:all}
\begin{eqnarray}
\chi_0&=&2.3(3)\times10^{-4}\,{\rm cm^{3}/mol}~,\nonumber\\
g&=&0.790(3)~,~~T_{\rm K}=97.8(6)\,{\rm K}~~({\cal J}=2)~~;\\
\label{CoqblinFitResults:a}
\nonumber\\
\chi_0&=&6.9(9)\times10^{-4}\,{\rm cm^{3}/mol}~,\nonumber\\
g&=&0.591(7)~,~~T_{\rm K}=103(2)\,{\rm K}~~({\cal J}=5/2)~~.
\label{CoqblinFitResults:b}
\end{eqnarray}
\end{mathletters}
The ${\cal J}=2$ curve fits our $\chi(T)$ data fairly well.  However, the 1.5
$d$-electrons per V ion could not give rise to a ${\cal J}$ value this large; the
very small value of $g$ is also considered highly unlikely.  

On the basis of the above analysis we conclude that the Coqblin-Schrieffer model
for $S > 1/2$ and the $S= 1/2$ Kondo model cannot explain the intrinsic
susceptibility of LiV$_2$O$_4$ over any appreciable temperature range.
\noindent

\section{Summary and Discussion}\label{ConclusionSec}

In this paper we have described the synthesis and characterization of nine
LiV$_2$O$_4$ samples.  Our magnetically purest samples 1 and 6 clearly showed a
broad shallow maximum in the observed magnetic susceptibility $\chi^{\rm obs}(T)$
at $T\approx 16$\,K, with small Curie-like upturns below $\sim 5$\,K\@. 
Field-cooled and zero-field-cooled magnetization measurements with
$H=10\mbox{--}100$\,G did not reveal any evidence for static spin-glass ordering
from 1.8--2 to 50\,K in any of the seven samples measured.  At $T\gtrsim 50$\,K,
$\chi^{\rm obs}(T)$ showed local magnetic moment behavior for all samples.  In
sample 2 which showed a larger Curie-like upturn in $\chi^{\rm obs}(T)$ at low
$T$ than in samples~1 and 6, we found that liquid-nitrogen quenching reduced the
Curie-like upturn to a large extent, revealing the broad peak in $\chi^{\rm
obs}(T)$.  However, ice-water quenching and slow-oven cooling enhanced the upturn,
and the above successful reduction of the upturn by liquid-nitrogen quenching
could not be reproduced.  We analyzed low-$T$ isothermal magnetization versus
applied magnetic field $M^{\rm obs}(H)$ data, and determined the parameters of the
paramagnetic impurities giving rise to the Curie-like upturn in $\chi^{\rm
obs}(T)$, assuming that a single type of impurity is present.  Using these
parameters, the intrinsic susceptibility $\chi(T)$ was obtained and found to be
essentially the same in all samples but one (4A).  Surprisingly, the spin $S_{\rm
imp}$ of the paramagnetic impurities was found to be large, $S_{\rm imp}=3/2$ to
4 depending on the sample, suggesting the presence of variable amounts of
ferromagnetically coupled vanadium spin defect clusters of variable size in the
samples.

We tested the localized magnetic moment picture for $\chi(T)$ at $T\gtrsim 50$\,K
using the HTSE prediction for the spin susceptibility of the $S=1/2$ vanadium
sublattice of the spinel structure, which yielded $C$ and $\theta$ values similar
to those reported in the past for LiV$_2$O$_4$.  Using the values of the Van Vleck
susceptibility obtained from $K$-$\chi$ analyses, the Pauli susceptibility
contribution to the temperature-independent susceptibility $\chi_0$ was derived
and found to be small, comparable to that of LiTi$_2$O$_4$.  The Van Vleck
formulas for the paramagnetic susceptibility of isolated V$^{+3}$ or V$^{+4}$ ions
or an equal mixture, assuming that each V ion is in a cubic CEF, failed to
describe the $T$ dependence of the observed effective magnetic moment.  For the
high-$T$ ``localized moment'' region, the observed effective moment is in
agreement with the spin-only value expected for $g\simeq 2$.

Our attempts to describe the low-$T$ susceptibility data in terms of the
single-ion Kondo ($S=1/2$) and Coqblin-Schrieffer (${\cal J}$ or $S > 1/2$) models
for isolated magnetic impurities in metals were unsuccessful.  These models
predict that the electronic specific heat coefficient $\gamma(T)$ and the
susceptibility $\chi(T)$ both show maxima for ${\cal J} \geq
3/2$.\cite{Rajan1983}  LiV$_2$O$_4$ clearly shows a peak in $\chi(T)$ at $T
\approx 16$\,K, but there is no peak in $\gamma(T)$ down to
1.2\,K.\cite{Kondo1997,Johnston1998}  Thus, these theories cannot
self-consistently explain the results of both measurements, suggesting that there
is some other mechanism responsible for the heavy-fermion behavior and/or that the
single-ion picture is inappropriate.  It is however intriguing that the
experimental Wilson ratio $R_{\rm W}\approx 1.7$ at 1\,K
(Ref.~\onlinecite{Kondo1997}) is close to that ($R_{\rm W}=2$) predicted for the
$S=1/2$ Kondo model.

In conventional $f$-electron heavy fermion compounds, local $f$-electron orbitals
and conduction electron states in non-$f$ bands hybridize only weakly, resulting
in a many-body scattering resonance of the quasiparticles near the Fermi energy
$E_{\rm F}$, a large density of quasiparticle states ${\cal D}(E_{\rm F})$, and
hence a large quasiparticle effective mass, electronic specific heat coefficient
and magnetic spin susceptibility at low $T$.  Screening of $S=1/2$ local moments
by conduction electron spins leads to a nonmagnetic ground state and a saturating
spin susceptibility as $T\rightarrow 0$.  In Sec.~\ref{ModelSec}, we tested
several models for $\chi(T)$ which assume the presence of local magnetic moments
in LiV$_2$O$_4$ which interact weakly with the conduction electrons.  However, in
these models as applied to LiV$_2$O$_4$, the itinerant and ``localized'' electrons
must both occupy $t_{2g}$ orbitals (or bands derived from these orbitals), rather
than orbitals of more distinct character.  One can imagine a scenario in which the
HF behaviors of LiV$_2$O$_4$ at low $T$ arise in a way similar to that of the
$f$-electron HF compounds, if the following conditions are fulfilled: (i) the
trigonal component of the CEF causes the $A_{1g}$ orbital singlet to lie below the
$E_{g}$ orbital doublet; (ii) one of the 1.5 $d$-electrons/V is localized in the
ground $A_{1g}$ orbital due to electron-electron correlations;\cite{Goodenough1}
(iii) the remaining 0.5 $d$-electron/V occupies the $E_{g}$ doublet and is
responsible for the metallic character; and (iv) the band(s) formed from the
$E_{g}$ orbitals hybridize only weakly with the $A_{1g}$ orbital on each V ion. 
This scenario involves a kind of orbital ordering; a more general discussion of
orbital ordering effects is given below.

The geometric frustration for antiferromagnetic ordering inherent in the V
sublattice of LiV$_2$O$_4$ may be important to the mechanism for the observed HF
behaviors of this compound at low $T$.  Such frustration inhibits long-range
magnetic ordering and enhances quantum spin fluctuations and (short-range)
dynamical spin ordering.\cite{Anderson1956,Villain1980,Canals1998}  These effects
have been verified to occur in the C15 fcc Laves phase intermetallic compound
(Y$_{0.97}$Sc$_{0.03}$)Mn$_2$, in which the Y and Sc atoms are nonmagnetic and the
Mn atom substructure is identical with that of V in LiV$_2$O$_4$.  In
(Y$_{0.97}$Sc$_{0.03}$)Mn$_2$, Shiga {\em et~al.} discovered quantum magnetic
moment fluctuations with a large amplitude ($\mu_{\rm rms}=1.3\,\mu_{\rm B}$/Mn at
8\,K) in their polarized neutron scattering study.\cite{Shiga1988}  They also
observed a thermally-induced contribution, with
$\mu_{\rm rms}=1.6\,\mu_{\rm B}$/Mn at 330\,K\@.  Further, Ballou {\em
et~al.}\cite{Ballou1996} inferred from their inelastic neutron scattering
experiments the presence of ``short-lived 4-site collective spin singlets,''
thereby suggesting the possibility of a quantum spin-liquid ground state.  A
recent theoretical study by Canals and Lacroix\cite{Canals1998} by perturbative
expansions and exact diagonalization of small clusters of a $S=1/2$ (frustrated)
pyrochlore antiferromagnet\cite{Gaulin1992} found a spin-liquid ground state and
an AF spin correlation length of less than one interatomic distance at $T=0$. 
Hence, it is of great interest to carry out neutron scattering measurements on
LiV$_2$O$_4$ to test for similarities and differences in the spin excitation
properties to those of (Y$_{0.97}$Sc$_{0.03}$)Mn$_2$.

(Y$_{0.97}$Sc$_{0.03}$)Mn$_2$ has some similarities in properties to those of
LiV$_2$O$_4$.  No magnetic long-range ordering was observed above 1.4\,K
(Refs.~\onlinecite{Ballou1996,Shiga1988}) and 0.02\,K,\cite{Kondo1997}
respectively.   Similar to LiV$_2$O$_4$, (Y$_{0.97}$Sc$_{0.03}$)Mn$_2$ shows a
large electronic specific heat coefficient $\gamma(0) \approx
160\mbox{--}200$\,mJ/mol\,K$^2$.\cite{Ballou1996,Wada1989}  However, the $T$
dependences of the susceptibility\cite{Nakamura1988} and $\gamma$
(Ref.~\onlinecite{Wada1989}) are very different from those seen in LiV$_2$O$_4$
and in the heaviest $f$-electron heavy fermion compounds.  $\chi^{\rm obs}(T)$
does not show a Curie-Weiss-like behavior at high $T$, but rather increases with
increasing $T$.\cite{Nakamura1988}  $\gamma(T)$ is nearly independent of $T$ up to
at least 6.5\,K\@.\cite{Wada1989}  Replacing a small amount of Mn with Al, Shiga
{\em et~al.}\ found spin-glass ordering in
(Y$_{0.95}$Sc$_{0.05}$)(Mn$_{1-x}$Al$_{x}$)$_2$ with
$x\geq 0.05$.\cite{Shiga1993}  The susceptibility for $x=0.15$ shows a
Curie-Weiss-like behavior above $\sim 50$\,K\@.  The partial removal of the
geometric frustration upon substitution of Al for Mn might be anologous to that
in our sample 3 in which structural defects evidently ameliorate the frustrated
V-V interactions, leading to spin-glass ordering below $\sim
0.8$\,K\@.\cite{Kondo1997}

The magnetic properties of materials can be greatly influenced when the ground
state has orbital degeneracy in a high-symmetry structure.  Such degenerate ground
state orbitals can become energetically unstable upon cooling.  The crystal
structure is then deformed to a lower symmetry to achieve a lower-energy,
non-orbitally-degenerate ground state (Jahn-Teller theorem).\cite{Kugel1973}  This
kind of static orbital ordering accompanied by a structural distortion is called
the cooperative Jahn-Teller effect.\cite{Kugel1973}  The driving force for this
effect is the competition between the CEF and the lattice energies.  Orbital
ordering may also be caused  by spin exchange interactions in a magnetic system
with an orbitally-degenerate ground state.\cite{Kugel1973,Pen1997}  The orbital
(and charge) degrees of freedom may couple with those of the spins in such a way
that certain occupied orbitals become energetically favorable, and consequently
the degeneracy is lifted.  As a result, the exchange interaction becomes spatially
anisotropic.  For example, Pen {\em et~al.}\cite{Pen1997} showed that the
degenerate ground states in the geometrically frustrated V triangular lattice
Heisenberg antiferromagnet LiVO$_2$ can be lifted by a certain static orbital
ordering.  X-ray and neutron diffraction measurements detected no structural
distortions or phase transitions in LiV$_2$O$_4$.\cite{Kondo1997,Chmaissem1997} 
However, the presence of orbital degeneracy or near-degeneracy suggests that
dynamical orbital-charge-spin correlations may be important to the physical
properties of LiV$_2$O$_4$.  It is not yet known theoretically whether such
dynamical correlations can lead to a HF ground state and this scenario deserves
further study.

Thus far we and collaborators have experimentally demonstrated heavy fermion
behaviors of LiV$_2$O$_4$ characteristic of the heaviest-mass $f$-electron HF
systems from magnetization,\cite{Kondo1997} specific
heat,\cite{Kondo1997,Johnston1998} nuclear magnetic
resonance,\cite{Kondo1997,Mahajan1998} thermal
expansion,\cite{Chmaissem1997,Johnston1998} and muon spin
relaxation\cite{Kondo1997} measurements.  Our magnetization study reported in
this paper was done with high-purity polycrystalline samples from which we have
determined the low temperature intrinsic susceptibility.  Nevertheless,
high-quality single crystals are desirable to further clarify the physical
properties.  In particular, it is crucial to measure the low-$T$ resistivity, the
carrier concentration and the Fermi surface.  In addition, when large crystals
become available, inelastic neutron scattering experiments on them will be vital
for a deeper understanding of this $d$-electron heavy fermion compound.

On the theoretical side, new physics may be necessary to explain the heavy fermion
behaviors we observe in LiV$_2$O$_4$.  We speculate that the geometric frustration
for antiferromagnetic ordering and/or coupled dynamical orbital-charge-spin
correlations may contribute to a new mechanism leading to a heavy fermion ground
state.  A successful theoretical framework must in any case self-consistently
explain the radically different properties of LiV$_2$O$_4$ and the isostructural
superconductor LiTi$_2$O$_4$.

\acknowledgements We are grateful to F.~Izumi for his help with our Rietveld
analyses using his \mbox{RIETAN-97$\beta$} program,\cite{Izumi1993} and to
Y.~Ueda, N.~Fujiwara and H.~Yasuoka for communications about their ongoing work
on LiV$_2$O$_4$.  We are indebted to A.~Jerez and N.~Andrei for providing their
Bethe ansatz calculation results for the $S=1/2$ Kondo
model.\cite{JerezAndrei1997}  We thank F.~Borsa, J.~B.~Goodenough, A.~V.~Mahajan,
R.~Sala, E.~Lee, I.~Inoue and H.~Eisaki for helpful discussions.  Ames Laboratory
is operated for the U.S. Department of Energy by Iowa State University under
Contract No.\ W-7405-Eng-82.  This work was supported by the Director for Energy
Research, Office of Basic Energy Sciences.

\end{document}